# NUMERICAL MODELLING OF WIND WAVES.
# PROBLEMS, SOLUTIONS, VERIFICATIONS, AND APPLICATIONS

V. G. Polnikov[1]

CONTENTS

[1] Research professor, A.M. Obukhov Institute for Physics of Atmosphere of Russian Academy of Sciences, Moscow, Russia 119017, e-mail: polnikov@mail.ru

# ABSTRACT

Due to stochastic feature of a wind-wave field, the time-space evolution of the field is described by the transport equation for the 2-dimensional wave energy spectrum density, $S(\sigma, \theta; \mathbf{x}, t)$, spread in the space, $\mathbf{x}$, and time, $t$. This equation has the forcing named the source function, $F$, depending on both the wave spectrum, $S$, and the external wave-making factors: local wind, W($\mathbf{x}$, $t$), and local current, U($\mathbf{x}$, $t$). The source function, $F$, is the "heart" of any numerical wind wave model, as far as it contains certain physical mechanisms responsible for a wave spectrum evolution.

It is used to distinguish three terms in function $F$: the wind-wave energy exchange mechanism, *In*; the energy conservative mechanism of nonlinear wave-wave interactions, *Nl*; and the wave energy loss mechanism, *Dis*, related, mainly, to the wave breaking and interaction of waves with the turbulence of water upper layer and with the bottom. Differences in mathematical representation of the source function terms determine general differences between wave models.

The problem is to derive analytical representations for the source function terms said above from the fundamental wave equations.

Basing on publications of numerous authors and on the last two decades studies of the author, the optimized versions of the all principal terms for the source function, $F$, have been constructed. Detailed description of these results is presented in this chapter.

The final version of the source function is tested in academic test tasks and verified by implementing it into numerical shells of the well known wind wave models: WAM and WAVEWATCH. Procedures of testing and verification are presented and described in details. The superiority of the proposed new source function in accuracy and speed of calculations is shown.

Finally, the directions of future developments in this topic are proposed, and some possible applications of numerical wind wave models are shown, aimed to study both the wind wave physics and global wind-wave variability at the climate scale, including mechanical energy exchange between wind, waves, and upper water layer.

1. INTRODUCTION

This chapter deals with theoretical description of wind wave phenomenon taking place at the air-sea interface. Herewith, the main aim of this description is directed to numerical simulation of the wind wave field evolution in space and time.

As an introduction to the problem, consider a typical scheme of the air-sea interface. In simplified approach it consists of three items (Fig. 1):

• Turbulent air boundary layer with the shear mean wind flow having a velocity value $\mathbf{W}_{10}(\mathbf{x})$ at the fixed horizon z =10m;

• Wavy water surface;

• Thing water upper layer where the turbulent motions and mean shear currents are present.

The main source of all mechanical motions of different space-time scales at the air-sea interface is a mean wind flow above the surface, which has variability scales of the order of thousand meters and thousand seconds. The turbulent part of a near-water layer (boundary layer) has scales smaller than a meter and a second. Variability of the wavy surface has scales of tens meters and ten seconds, whilst the upper water motions have a wide range of scales covering all mentioned values. Thus, the wind impacts on the water upper layer indirectly via the middle scale motions of wind waves, and this impact is spread through a wide range of scales, providing the great importance of wind wave motion on the global scale.

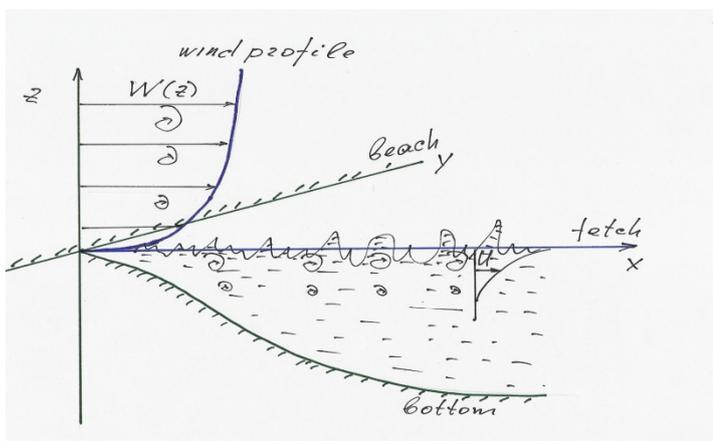

Fig. 1. The air-sea interface system

Besides of the said, this phenomenon has its own scientific and practical interest. The former is provided by a physical complexity of the system, whilst the latter is due to dangerous feature of the phenomenon. All these features justify the long period interest to the problem of wind wave modeling, staring from the well know paper by Stokes (1847).

From scientific point of view it is important to describe in a clear mathematical form a whole system of mechanical interactions between items mentioned above, responsible for the exchange processes at the air-sea interface. This is the main aim of the interface hydrodynamics. From practical point of view, a mathematical description of these processes permits to solve a lot of certain problems. As an example of such problems one may point out an improvement of wave and wind forecasting, calculation of heat and gas exchange between atmosphere and ocean, surface pollution mixing and diffusion, and so on.

Direct mathematical description of mechanical exchange processes in the system considered is very complicated due to multi-scale and stochastic nature of them (for example, see

Kitaigorodskii & Lamly, 1983). It can not be done in an exact form. Nevertheless, real advantage in this point can be reached by consideration of the problem in a spectral representation. Up to the date, a principal physical understanding exchange processes at the air-sea interface was achieved to some extent (Proceedings of the symposium on the wind driven air-sea interface, 1994; 1999), and mathematical tool for their description in spectral representation was constructed (for example, see Hasselmann, 1962; Zakharov, 1974; Phillips, 1977). Thus, one may try to make description of main processes at the air-sea interface from the united point of view. Below, we consider the main theoretical procedures needed to manage this problem.

## 2. FUNDAMENTAL EQUATIONS AND CONCEPTIONS

From mathematical point of view, a wind wave field is a stochastic dynamical process, and the properties of this field should be governed by a proper statistical ensemble. Therefore, the best way of the phenomenon description lies in the domain of statistical characteristics, the main of which for a non-stationary field is the two-dimensional spatial wave energy spectrum, $S(\mathbf{k},\mathbf{x},t) \equiv S$, spread in the space, $\mathbf{x}$, and time, $t$.

Traditionally, the space-time evolution of this characteristic is described by the so called transport equation written in the following spectral representation (Komen et al, 1994)

$$\frac{\partial S}{\partial t} + C_{gx}\frac{\partial S}{\partial x} + C_{gy}\frac{\partial S}{\partial y} = F \equiv Nl + In - Dis. \tag{2.1}$$

Here, the left-hand side is the full time-derivative of the spectrum, and the right-hand side is the so called source function ("forcing"), $F$. Vector ($C_{gx}$, $C_{gy}$) is the group velocity one, corresponding to a wave component with wave vector $\mathbf{k}$, which is defined by the ratio

$$\mathbf{C}_g = \frac{\partial \sigma(k)}{\partial k}\frac{\mathbf{k}}{k} = (C_{gx}, C_{gy}). \tag{2.2}$$

Dependence of frequency $\sigma(\mathbf{k})$ on the wave vector $\mathbf{k}$ is given by the expression

$$\sigma = \sqrt{gk}, \tag{2.3}$$

known as the dispersion relation for the case of deep water, considered below.

The left-hand side of equation (2.1) is responsible for the "mathematical" part of model. The physical essence of model is held by the source function, $F$, depending on both the wave spectrum, $S$, and the external wave-making factors: local wind, W($\mathbf{x}$,$t$), and local current, U($\mathbf{x}$,$t$).

At present, it is widely recognized that $F$ can be written as a sum of three terms – three parts of the united evolution mechanism for wind waves:

- The rate of conservative nonlinear energy transfer through a wave spectrum, $Nl$, ("nonlinear-term");

- The rate of energy transfer from wind to waves, $In$, ("input-term");

- The rate of wave energy loss due to numerous dissipative processes, $Dis$, ("dissipation-term").

The source function is the "heart" of the model. It describes certain physical processes included in the model representation, which determine mechanisms responsible for the wave spectrum evolution (Efimov& Polnikov, 1991; Komen et al, 1994). Differences in representation of the source function terms mentioned above determine general differences between different wave models. In particular, the models are classified with the category of generations, by means of ranging the parameterization for *Nl*-term (The SWAMP group, 1985). This classification could be extended, taking into account all source function terms (for example, see Polnikov, 2005, 2009; Polnikov&Tkalich, 2006). The worldwide spread models WAM (The WAMDI group, 1988) and WAWEWATCH (WW) (Tolman&Chalikov, 1996) are the representatives of such a kind models, which are classified as the third generation ones.

Differences in representation of the left hand side of evolution equation (2.1) and in realization of its numerical solution are mainly related to the mathematics of the wave model. Such a kind representation determines specificity of the model as well. But it is mainly related to the category of variation the applicability range of the models (i.e. accounting for a sphericity of the Earth, wave refraction on the bottom or current inhomogeneity, and so on). We will not dwell on this issue more in this chapter.

Note that equation (2.1) has a meaning of the energy conservation law applied to each spectral component of wave field. Nevertheless, to have any physical meaning, this equation should be derived from the principal physical equations. By this way, the most general expressions for the source terms could be found. And this is the main problem of the task considered here.

Since pioneering paper by Stokes (1847), the basic hydrodynamic equations, describing the wave dynamics at the interface of an ideal liquid, are as follows

$$\rho \frac{d\mathbf{u}}{dt} = -\vec{\nabla}_3 P - \rho \mathbf{g} + \mathbf{f}(\mathbf{x},t); \big|_{z=\eta(\mathbf{x},t)} \quad , \tag{2.4}$$

$$\frac{\partial \rho}{\partial t} + \vec{\nabla}_3 (\rho \mathbf{u}) = 0 \quad , \tag{2.5}$$

$$u_z \big|_{z=\eta(\mathbf{x},t)} = \frac{\partial \eta}{\partial t} + (\mathbf{u}\vec{\nabla}_2 \eta) \quad , \tag{2.6}$$

$$u_z \big|_{z=-\infty} = 0 \quad . \tag{2.7}$$

Here, the following designations are used:

$\rho(z,t)$ is the fluid density;

$\mathbf{u}(\mathbf{x},z,t) = (u_x, u_y, u_z)$ is the velocity field;

$P(\mathbf{x},z,t)$ is the atmospheric pressure;

$g$ is the acceleration due to gravity;

$f(\mathbf{x},z,t)$ is the external forcing (viscosity, surface tension, wind stress and so on);

$\eta(\mathbf{x},t)$ is the surface elevation field;

$\mathbf{x} = (x, y)$ is the horizontal coordinates vector;

$z$ is the vertical coordinate up-directed;

$\vec{\nabla}_2 = (\frac{\partial}{\partial x}, \frac{\partial}{\partial y})$ is the horizontal gradient vector;

$\vec{\nabla}_3 = (\vec{\nabla}_2, \frac{\partial}{\partial z})$ is the full gradient,

and the full time-derivative operator is defined as $\frac{d}{dt}(...) = \left(\frac{\partial}{\partial t} + \mathbf{u}\vec{\nabla}_3\right)(...)$.

We remind that Eq. (2.4) is the main dynamic equation used at the water surface $z = \eta(x,t)$, Eq. (2.5) is the mass conservation law, Eq. (2.6) is the kinematical boundary condition at surface $\eta(x,t)$, and Eq. (2.7) is the boundary condition at the bottom. Note that Eqs. (2.4) and (2.6) are principally nonlinear.

General problem is to derive all source terms from the set of equations (2.4)-(2.7), taking into account a stochastic feature for motions near interface. It is easy to understand that the posed problem is quite complicated. Nevertheless, it can be solved under some approximations, if one takes into account each evolution mechanism separately. The history of such investigations is described in quite numerous papers, the main results of which are accumulated in numerous books (Komen et al, 1994; Young, 1999; and others). Below we reconstruct some principal results of these papers, permitting us to show the state-of-the-art in this field of hydrophysics.

To this end, first of all, one should introduce the rules of transition form physical fields variables, $\mathbf{u}(\mathbf{x},z,t)$, $\eta(\mathbf{x},t)$, and $f(\mathbf{x},z,t)$, to its spectral representation. To do this, the so called Fourier-Stiltjes decomposition is introduced for each of the fields mentioned. As far as the main equations are used at the interface surface, we demonstrate this decomposition procedure on the example of surface elevation field. In such a case, one writes

$$\eta(\mathbf{x},t) = const \cdot \int_{\mathbf{k}} \exp[i(\mathbf{k}\mathbf{x})]\eta_{\mathbf{k}}(t)d\mathbf{k} \qquad . \qquad (2.8)$$

Here, $\eta_{\mathbf{k}}(t)$ is the so called Fourier-amplitude of the field $\eta(\mathbf{x},t)$, taking in mind that this field is non-stationary, but homogeneous. In such a case, only, the exponential decomposition is effective in a further simplification of the equations (for details, see, Monin&Yaglom, 1971). By substitution of the decompositions of the kind (2.8) for each field into the system of Eqs. (2.4)-(2.7), one could get the final equation for the main variable, $\eta_{\mathbf{k}}(t)$, in the form

$$\partial \eta_{\mathbf{k}} / \partial t = func1[\eta_{\mathbf{k}}, u_{\mathbf{k}}, f_{\mathbf{k}}] \qquad (2.9)$$

where the right hand side of (2.9) represents a complicated functional having as an arguments the Fourier-amplitudes for each field variables.

Then, one introduces the wave energy spectrum $S(\mathbf{k})$ by the rule

$$<< \eta_{\mathbf{k}}^{s}(t)\eta_{\mathbf{k}'}^{s'}(t) >> = S(\mathbf{k},t)\delta(\mathbf{k}-\mathbf{k}')\delta(s+s'). \qquad (2.10)$$

Here, the brackets $<<...>>$ means the wave statistical ensemble averaging, indexes $s$ and $s'$ means the sign of imaginary part of the complex conjugated form for $\eta_{\mathbf{k}}(t)$, and $\delta(...)$ is the Dirak's delta-function.

After this, one should execute some procedure, including multiplication of Eq. (2.9) by the complex conjugated amplitude, $\eta_{\mathbf{k}}^{*}(t)$, and summation the resulting equation with the complex conjugated one. Finally, one gets the final equation of the form

$$\partial S(\mathbf{k}) / \partial t = func2[\eta_{\mathbf{k}}, \eta_{\mathbf{k}}^{*}, u_{\mathbf{k}}, f_{\mathbf{k}}] \qquad , \qquad (2.11)$$

resembling Eq. (2.1), but written without spatial derivatives (i.e. for a homogeneous field). The right hand side of Eq. (2.11), if specified, is the sought source function.

Thus, with some simplifications and introducing certain rules of statistical averaging, the theoretical way of derivation for the spectrum evolution equation is found. Generalization of the procedure described above for the case of inhomogeneous wave field, $\eta(\mathbf{x},t)$, is less important for our further consideration. The readers could find this generalization in paper (Zaslavskii& Lavrenov, 2005). So, we could start to consider description and parameterizations for the definite source function terms responsible for certain physical evolution mechanisms of wind wave spectrum, which are needed further for a wind wave numerical simulation.

## 3. WAVE EVOLUTION MECHANISM DUE TO NONLINEARITY

*3.1. General grounds*

Nonlinear feature of fundamental equations (4) and (6) leads to a certain nonlinear mechanism of evolution for wave spectrum. This mechanism, described by the source term *Nl*, is theoretically the most investigated among others evolution mechanisms. For the first time the *Nl* term was derived by Hasselmann (1962), under some reasonable theoretical suggestions. Later it was rederived in the terms of very elegant Hamiltonian formalism by Zakharov (1968, 1974). Following to the latter papers, we show several points of this derivation.

First of all, we point out that the nonlinear theory is constructed under very certain suggestions: potential motions, no external forcing, and weak nonlinearity. The latter means existence of small values for wave steepness parameter defined by

$$\varepsilon = k_p <<\eta^2>>^{1/2} \quad (3.1)$$

where $k_p$ is the modulus of the dominant wave vector corresponding to the peak of spectrum, and $<<\eta^2>>^{1/2}$ is the mean wave amplitude.

For the ideal fluid without forcing and the potential motion approximation, it was shown (Zakharov, 1968) that the system of Eqs. (2.4)-(2.7) is the Hamiltonian one. It means that the system can be replaces by classical Hamiltonian equations of the form

$$\frac{\partial \eta}{\partial t} = \frac{\delta H}{\delta \Phi}, \quad (3.2)$$

$$\frac{\partial \Phi}{\partial t} = -\frac{\delta H}{\delta \eta} \quad (3.3)$$

where $H(\eta,\Phi)$ is the Hamiltonian function of two canonical variables. Zakharov has found that canonical variables are the surface elevation, $\eta(\mathbf{x},t)$, and the velocity potential at this surface, $\Phi(\mathbf{x},t) \equiv \varphi(\mathbf{x},z,t)|_{z=\eta(\mathbf{x},t)}$, where $\vec{\nabla}_3 \varphi(\mathbf{x},z,t) = \mathbf{u}(\mathbf{x},z,t)$ is the wave velocity potential. Eventually, in terms of Hamiltonian formalism, the system of nonlinear equations (2.4)-(2.7) can be reduced to one of the form

$$i\frac{\partial a(\mathbf{k})}{\partial t} = \sigma(\mathbf{k})a(\mathbf{k}) +$$

$$+ \int \left[ U^{(1)}_{0,1,2} a(\mathbf{k}_1)a(\mathbf{k}_2)\delta_{0-1-2} + 2U^{(1)}_{1,0,2} a(\mathbf{k}_1)a^*(\mathbf{k}_2)\delta_{1-0-2} + U^{(3)}_{0,1,2} a^*(\mathbf{k}_1)a^*(\mathbf{k}_2)\delta_{0+1+2} \right] d\mathbf{k}_1 d\mathbf{k}_2 +$$

$$+ \int V^{(2)}_{0,1,2,3} a^*(\mathbf{k}_1)a(\mathbf{k}_2)a(\mathbf{k}_3)\delta_{0+1-2-3} d\mathbf{k}_1 d\mathbf{k}_2 d\mathbf{k}_3 \quad (3.4)$$

where $U^{(n)}_{0,1,2}$, $V^{(n)}_{0,1,2,3}$ are known functions of $\mathbf{k}_i$ ($i = 0,1,2,3$), and $\delta_{0-1-2} = \delta(\mathbf{k}-\mathbf{k}_1-\mathbf{k}_2)$ are the Dirak's delta-functions.

For us it is important that equation (3.4) is an analog of eq. (2.9) with the only difference that Fourier-amplitude $a(\mathbf{k})$ is the specially defined so called "normal" variable related linearly to the Fourier-amplitudes for elevation, $\eta_\mathbf{k}$, and for wave velocity potential at the surface, $\Phi_\mathbf{k}$. The spectrum defined for amplitude $a(\mathbf{k})$ by the rule (2.10) is called as the wave action spectrum, $N(\mathbf{k})$, which is linearly related to the wave energy spectrum, $S(\mathbf{k})$, by definition of variable $a(\mathbf{k})$. Details of this theory are not principal for us here. The most important fact is that the standard procedure described above results in the final evolution equation for spectrum $N(\mathbf{k})$ with the right hand side provided by the nonlinearity of the system under consideration. Just this part is the nonlinear evolution term, $Nl(\mathbf{k})$.

At present, it is well recognized that the exact theoretical result for the nonlinear source term has the kind of the so called four-wave kinetic integral

$$\partial N(\mathbf{k}_4)/\partial t \equiv Nl[N(\mathbf{k}_4)] = 4\pi \int d\mathbf{k}_1 \int d\mathbf{k}_2 \int d\mathbf{k}_3 M^2(\mathbf{k}_1, \mathbf{k}_2, \mathbf{k}_3, \mathbf{k}_4) \times$$
$$[N(\mathbf{k}_1)N(\mathbf{k}_2)(N(\mathbf{k}_3) + N(\mathbf{k}_4)) - N(\mathbf{k}_3)N(\mathbf{k}_4)(N(\mathbf{k}_1) + N(\mathbf{k}_2))] \times$$
$$\times \delta[\sigma(k_1) + \sigma(k_2) - \sigma(k_3) - \sigma(k_4)]\delta(\mathbf{k}_1 + \mathbf{k}_2 - \mathbf{k}_3 - \mathbf{k}_4) \quad . \quad (3.5)$$

Here, $\mathbf{k}_i$ is the wave vector having to a proper frequency-angular wave component $(\sigma_i, \theta_i)$ ($i = 1,2,3,4$), $M(\ldots)$ are the matrix elements describing intensity of interactions between four waves, and $\delta(\ldots)$ is the Dirak's delta-function providing the resonance feature of interactions.

The properties of kinetic integral (3.5) were studied numerically in numerous papers for very different spectral shapes (for references, see Polnikov&Farina, 2002; or, in more details, Polnikov, 2007). It was found that the main feature of $Nl(\sigma,\theta)$ is a permanent shift of the spectrum to lower frequencies with a very certain changing the shape of spectrum (Fig. 2).

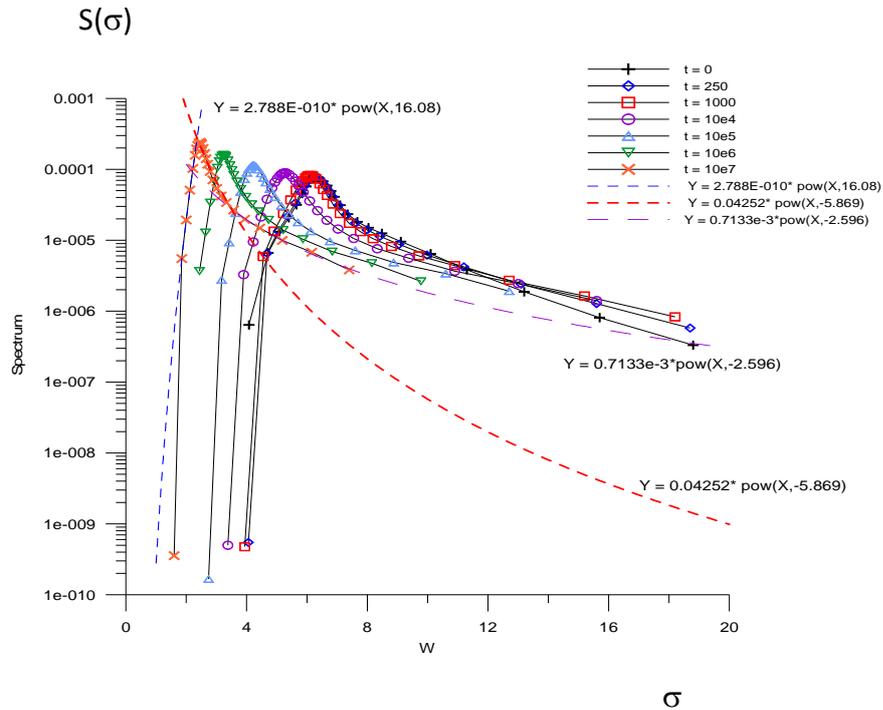

Fig. 2. Example of numerical solution of wave evolution due to nonlinear mechanism
(following to Lavrenov& Polnikov, 2001)

Additionally it should be noted that $Nl$-mechanism is proportional to the 6th power of small parameter of nonlinearity, $\varepsilon$. For this reason, the rate of nonlinear energy transfer through the spectrum is proportional to $\varepsilon^4$, i.e. it is very slow mechanism. Nevertheless, just $Nl$ term is responsible for wave length enlarging in the course of wind wave evolution, permitting waves to grow to large heights, but maintaining small value of wave steepness, $\varepsilon$.

Regarding to practical use of these theoretical results, due to great complicity of term (3.5), it needs to find an optimal approximation of the kinetic integral, which conserves all its properties.

This point was a subject of numerous investigations, among which there is a special research executed in Polnikov&Farina (2002) and Polnikov (2003). By comparing several theoretically justified approximations for $Nl(\sigma,\theta)$, it was found that the most efficient one (in terms of a specially defined criterion for accuracy and speed of calculation) is the discrete interaction approximation (DIA) proposed by Hasselmann et al. (1985). Just this approximation is used in widely spread models WAM and WAVEWATCH. Due to this, it is worth wile to dwell in more details on these results.

### 3.2. The most effective approximation for Nl term

First of all, let us say several words about DIA approximation.

This approximation means replacing the whole continuum of summands under kinetic integral (3.5) by the single one. This single summand is especially chosen, to get the best accuracy against exact kinetic integral, and the choice is defined by the geometry of four interacting wave vectors ($\mathbf{k}_i$ where $i = 1,2,3,4$). The optimal configuration of four wave vectors in DIA, proposed in original paper (Hasselmann et al, 1985) is shown in Fig. 3. In the polar coordinates, $(\sigma,\theta)$, the original configuration is governed by the following rations:

1) $\mathbf{k}_1 = \mathbf{k}_2 = \mathbf{k}$, where the arbitrary wave vector $\mathbf{k}$ is represented by $\sigma$ and $\theta$;

2) $\mathbf{k}_3 = \mathbf{k}_+$,   where $\mathbf{k}_+$ is represented by $\sigma_+ = \sigma(1+\lambda)$ and $\theta_+ = \theta + \Delta\theta_+$;

3) $\mathbf{k}_4 = \mathbf{k}_-$,   where $\mathbf{k}_-$ is represented by $\sigma_- = \sigma(1-\lambda)$ and $\theta_- = \theta - \Delta\theta_-$;

4) Parameters of the configuration are

$$\lambda = 0.25, \ \Delta\theta_+ = 11.5°, \text{ and } \Delta\theta_- = 33.6°. \tag{3.6}$$

In such a case, the nonlinear term $Nl(\mathbf{k})$ at all mentioned $\mathbf{k}$-points takes the form

$$Nl(\mathbf{k}_-) = I(\mathbf{k},\mathbf{k}_+,\mathbf{k}_-), \ \ Nl(\mathbf{k}_+) = I(\mathbf{k},\mathbf{k}_+,\mathbf{k}_-), \ \ Nl(\mathbf{k}) = -2I(\mathbf{k},\mathbf{k}_+,\mathbf{k}_-), \tag{3.7}$$

where

$$I(\mathbf{k},\mathbf{k}_+,\mathbf{k}_-) = C_{NL} g^{-8} \sigma^{19} \left[ N^2(\mathbf{k})\left(N(\mathbf{k}_+) + N(\mathbf{k}_-)\right) - 2N(\mathbf{k})N(\mathbf{k}_+)N(\mathbf{k}_-) \right]. \tag{3.8}$$

In (1.19) $C_{NL}$ is the fitting constant. The net nonlinear term at any fixed $(\sigma,\theta)$-point is found by the procedure of running of Eqs. (3.7)-(3.8) through all points of the frequency-angle integration grid $\{\sigma_i,\theta_j\}$.

According to study Polnikov&Farina (2002), features of the original DIA are as follows.

F1. A mean relative error (over the set of 16 representative spectral shapes) is about 60%.

F2. Relative consuming time for calculation of $Nl$ term (in WAM with DIA approximation) is about 48% of the CPU.

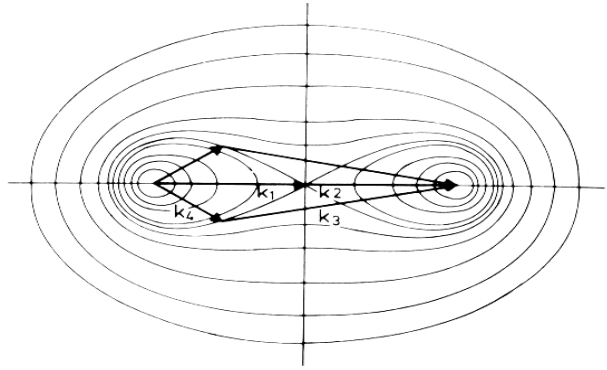

Fig. 3. The original optimal configuration of four wave vectors in DIA (Hasselmann et al, 1985)

Despite of such a kind features, in the study mentioned it was found that DIA is the best among others approximations. Herewith, a simple consideration has shown that the reason of small accuracy is due to replacing the exact kinetic integral by the only summand. The reason of high consuming time is provided by necessity to make the spectrum interpolation for wave vectors $\mathbf{k}_3$ and $\mathbf{k}_4$, which are not allocated at the frequency-angle integration grid $\{\sigma_i, \theta_j\}$.

In the studies mentioned it was found that the shortages said could be reduced to some extent.

The first step of approximation improvement was a proposal to refuse of the resonant feature of four-wave interaction, and to allocate all four vectors at the integration grid $\{\sigma_i, \theta_j\}$. In such a case, the new configuration was constructed, which is shown in Fig. 4 by dashed lines. (Moreover, the vectors $\mathbf{k}_1$ and $\mathbf{k}_2$ are meant to be allocated on the integration grid, but they are not necessarily equal each to other.) It allows us to exclude the interpolation procedure and make calculation of *Nl* faster. This version of DIA was named as FDIA. The relative consuming time in FDIA is reduced twice (improvement of feature F2).

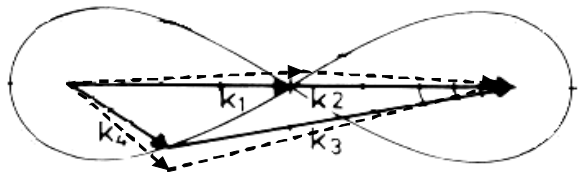

Fig. 4. The configuration of Fast DIA is shown by dashed lines.

The second step of DIA improvement was a choice of more efficient configuration. After some attempts, the best configuration was found (Polnikov&Farina, 2002; Polnikov, 2003). This has permitted us to reduce the relative error of approximation to 40% (improvement of feature F1).

Taking into account these two advantages, we propose to use the following new version of FDIA for a numerical representation of *Nl* term.

(1) The calculating frequency-angular grid, $\{\sigma_i, \theta_j\}$, is defined typically:

exponential frequency grid

$$\sigma(i) = \sigma_0 e^{i-1} \quad (1 \leq i \leq N), \tag{3.9a}$$

and equidistant angular grid

$$\theta(j) = -\pi + (j-1) \cdot \Delta\theta \quad (1 \leq j \leq M), \tag{3.9b}$$

where $\sigma_0$, $e$, and $\Delta\theta$ are the grid parameters specified in a certain numerical model.

(2) The reference wave component, $(\sigma_4, \theta_4)$, is allocated at any current node of the grid (1.20).

(3) The other 3 waves have the components allocated at nodes of the same grid and defined by the ratios

$$\sigma_1 = \sigma_4 e^{m1}, \quad \sigma_2 = \sigma_4 e^{m2}, \quad \sigma_3 = \sigma_4 e^{m3}, \tag{3.10a}$$

$$\Delta\theta_1 = n1\,\Delta\theta,\ \Delta\theta_2 = n2\,\Delta\theta,\ \Delta\theta_3 = n3\,\Delta\theta, \text{(where } \Delta\theta_i \equiv |\theta_i - \theta_4|). \tag{3.10b}$$

Thus, the optimal configuration of four interacting waves is given by the certain set of integers: $m1$, $m2$, $m3$; $n1$, $n2$, $n3$, which, in turn, are to be especially calculated, in dependence on the grid parameters, $e$ and $\Delta\theta$ (Polnikov&Farina, 2002).

For the frequency-angle grid with parameters $e = 1.1$ and $\Delta\theta = \pi/12$, which are typical for the models WAM and WW, the most effective configuration is given by the following parameters (Polnikov, 2003)

$$m1 = 3,\ m2 = 3,\ m3 = 5;\quad n1 = n2 = 2,\ n3 = 3. \tag{3.11}$$

Finally, $Nl$-term is calculated by the standard formulas (3.7)-(3.8), making the loops for the reference components, $(\sigma, \theta)$, arranged through the grid (3.9). For completeness, these formulas for the energy spectrum representation are as follows

$$Nl(\sigma_4,\theta_4) = Nl(\sigma_3,\theta_3) = I(\sigma_1,\theta_1,\sigma_2,\theta_2,\sigma_3,\theta_3,\sigma_4,\theta_4), \tag{3.12a}$$

$$Nl(\sigma_1,\theta_1) = Nl(\sigma_2,\theta_2) = -I(\sigma_1,\theta_1,\sigma_2,\theta_2,\sigma_3,\theta_3,\sigma_4,\theta_4), \tag{3.12b}$$

where

$$I(...) = C_{nl} g^{-4} \sigma_4^{11} \left[ S_1 S_2 (S_3 + (\sigma_3/\sigma_4)^4 S_4) - S_3 S_4 ((\sigma_2/\sigma_4)^4 S_1 + (\sigma_1/\sigma_4)^4 S_2) \right]. \tag{3.13}$$

Here, $C_{nl}$ is the only fitting dimensionless coefficient, and notation $S_i \equiv S(\sigma_i, \theta_i)$ is used. The coefficient $C_{nl}$ should be fitted together with other source terms simultaneously.

Hereby, the modern and comprehensive description of $Nl$ term parameterization is finished, and we can go to the next terms.

## 4. WIND WAVE ENERGY PUMPING MECHANISM

### *4.1. General grounds*

This mechanism is studied rather well both theoretically and empirically (for references, see recent review paper, The WISE group, 2007). Nevertheless, it is known much more less than *Nl*-mechanism described above in previous section 3. The reason of such a situation is a high complexity of processes taking place in a boundary layer of atmosphere located in the vicinity of waving surface. To get the main understanding of the problem, let us consider the simplest approach proposed in the earlier papers by Phillips (1957) and Miles (1960) and summarized in book (Phillips, 1977).

To the purpose said, let us rewrite the system of Eqs. (2.4)-(2.7) in a linear approximation and potential motion approach. We have

$$\frac{\partial \Phi}{\partial t} = -g\eta + P(a_p) \quad , \tag{4.1}$$

$$\frac{\partial \eta}{\partial t} = \frac{\partial \Phi}{\partial z} \quad , \tag{4.2}$$

$$\Delta \varphi = 0 \quad , \tag{4.3}$$

$$\left.\frac{\partial \varphi}{\partial z}\right|_{z=-\infty} = 0 \quad . \tag{4.4}$$

In this representation, the first two equations are written at the surface $\eta(\mathbf{x},t)$, whilst two equations, (4.3) and (4.4), done in the whole water layer. The last term in the r. h. s. of (4.1), $P(a_p)$, means the result of transition to the potential representation for the atmospheric pressure forcing, real and unequivocal representation of which is not specified yet. Here, $P(a_p)$ is normalized by the water density, $\rho_w$, and $a_p$ is the formal argument of the forcing function.

For more certainty, note that the first two equations, (4.1) and (4.2), are the main ones, as far as they determine unknown fields, $\eta(\mathbf{x},t)$ and $\Phi(\mathbf{x},t)$. Equations (4.3) and (4.4) are auxiliary ones; they are used for determination of the vertical structure for velocity potential $\varphi(\mathbf{x},z,t)$, only.

To make a transition into the spectral representation, one introduces the following Fourier-decompositions

$$\eta(\mathbf{x},t) = const \cdot \int_{\mathbf{k}} \exp[i(\mathbf{kx})]\eta_{\mathbf{k}}(t)d\mathbf{k}, \tag{4.5}$$

$$\varphi(\mathbf{x},z,t) = const \cdot \int_{\mathbf{k}} \exp[i(\mathbf{kx})]f(z)\varphi_{\mathbf{k}}(t)d\mathbf{k} \quad . \tag{4.6}$$

Hereafter, the wave vector, **k,** as well as the spatial vector, **x,** has the horizontal components, only, and *f(z)* is the so called vertical structure function, determined from Eqs. (4.3) and (4.4), solved analytically. In our case, *f(z)*=exp(-*kz*), what dose not play any principal role later. After substitution of representations (4.5), (4.6) into the system of Eqs. (4.1)-(4.4), equations (4.3) and (4.4) give the solution for function *f(z)* mentioned above, and the other two equations get the kind

$$\dot{\Phi}_{\mathbf{k}} + g\eta_{\mathbf{k}} = \Pi(\mathbf{k},a_p) \quad , \tag{4.7}$$

$$\dot{\eta}_{\mathbf{k}} = k\Phi_{\mathbf{k}} \quad . \tag{4.8}$$

The solution of system (4.7)-(4.8) radically depends on representation of the atmospheric pressure forcing, $\Pi(\mathbf{k}, a_p)$. This representation is the main theoretical problem which is not solved wholly till present.

There are some simple solutions of the system (4.7)-(4.8), resulting in two principle mechanisms of wave pumping: so called Philips' and Miles' mechanisms. Their simplified treatments are as follows (for details, see Phillips, 1977).

The atmospheric pressure forcing can be represented in the form

$$\Pi(\mathbf{k}, a_p) = (\alpha + i\beta)c^2 k \eta_k(t) + \gamma_k(t) \qquad (4.9)$$

where the first term in the right hand side means the part of pressure oscillations induced by waves and the second term means the pure turbulent oscillations uncorrelated with wave motions. Besides, the induced oscillations have two constituents: one of them, having intensity coefficient $\alpha$, is in phase with the surface elevations; and the other one, having intensity coefficient $\beta$, is in phase with the wave slope. As usually, $c = c(k) = \sigma(k)/k$ is the phase velocity of the wave component with vector $\mathbf{k}$.

Substitution of (4.9) into system (4.7)-(4.8) with some rather complicated mathematics (via getting an equation of the form 2.9) leads to the following general solution for the energy wave spectrum $S(\mathbf{k},t)$ defined by rule (2.10):

$$S(\mathbf{k}, t) = C_{in} A(\mathbf{k}, \sigma) \left( \frac{\text{sh}(\beta \sigma t)}{\beta \sigma} \right) \qquad (4.10)$$

Here, $C_{IN}$ is the fitting coefficient of the theory, and $A(\mathbf{k}, \sigma)$ is the 3-dimensional spectrum of the turbulent part of atmospheric oscillations, the kind of which it the main problem of the theory. But the temporal behavior of wave spectrum evolution is clear from result (4.10).

Really, on the small time scale, when $\beta \sigma t \ll 1$, wave spectrum is proportional to time

$$S(\mathbf{k}, t) = C_{in} A(\mathbf{k}, \sigma) t \qquad (4.11)$$

what is equivalent to the evolution equation

$$\partial S(\mathbf{k}, t) / \partial t = C_{in} A(\mathbf{k}, \sigma) . \qquad (4.12)$$

In such a case the sought input term, $In$, is independent of wave spectrum. Such mechanism was called as the Phillips' one (after paper Phillips, 1957). This mechanism is working on very small time scales, and for this reason it is not used in numerical simulation on oceanic scales. The role of result (4.11) is an explanation of the wind wave origin.

The opposite case, when $\beta \sigma t \gg 1$, solution (4.10) results in the exponential growth of wave spectrum, what is equivalent to evolution equation of the form

$$\partial S(\mathbf{k}, t) / \partial t = C_{in} \beta \sigma S(\mathbf{k}, t) = C_{in} \sigma \beta(\mathbf{k}, \mathbf{W}) S(\mathbf{k}, t) \qquad (4.13)$$

where the final expression of $In$ term is written in the commonly used form. Dimensionless function $\beta(\mathbf{k}, \mathbf{W})$, depending of wave vector and wind at the reference horizon, is often called as the Miles' increment coefficient (or function). In accordance to the said earlier, specification of this function is the main difficulty of the interface dynamic theory and experiment.

Do not dwelling on this problem in details, below we shall barely mention some experimental results (Snyder et al, 1981; Plant, 1982) and numerical simulation ones (Chalikov, 1980, 1998, 2002), which gave the grounds for practically used parameterizations of $\beta(\mathbf{k}, \mathbf{W})$.

*4.2. Effective approximations for In term*

Understanding the difficulty of measurements for wind-wave energy exchange is following from the fact that the first more or less accurate experimental results have been obtained in famous work by Snyder et al (1981), only, i.e. much later than the first theoretical results mentioned above. On the basis of their results, Snyder and coauthors have constructed the following simplest parameterization

$$\beta(\sigma, \theta, \mathbf{U}) = \max\left[0, a \frac{\rho_a}{\rho_b}\left(\frac{W_5 \sigma}{g}\cos(\theta - \theta_u) - b\right)\right], \tag{4.14}$$

where $\rho_a$ and $\rho_w$ is the air and water density, respectively, $g$ is the gravity acceleration, $a$ and $b$ are the fitting parameters, and $W_5$ and $\theta_u$ is the local wind at horizon z=5m and its direction, respectively. Parameters $a$ and $b$ are varying in the following intervals: $a \cong 0.2\text{-}0.3$ и $b \cong 0.9\text{-}1$. The main disadvantage of result (4.14) was a very short range of validity: $1 < W_5\sigma/g < 3$.

Only one year later, Plant (1982) have accumulated a lot of experimental data and proposed an alternative parameterization

$$\beta = (0.04 \pm 0.02)\left(\frac{u_* \sigma}{g}\right)^2 \cos(\theta - \theta_u) \tag{4.15}$$

where

$$u_* = C_d^{1/2}(z) W(z) \tag{4.16}$$

is the so called friction velocity, and $C_d(z)$ was the drag coefficient for the horizon $z$. Advantage of this result is a much wider range of validity:

$$2.5 < W_{10}\sigma/g < 75. \tag{4.17}$$

But the behavior of $\beta$ in the vicinity of $W_{10}\sigma/g \cong 1$ was remaining unclear.

This uncertainty was vanished in numerical simulation of air-sea interface dynamics, started by Chalikov (1980). By direct numerical solution of proper equations for wave and air motions in the interface boundary zone, he have found the rate of wind-wave energy exchange and proposed very effective parameterization of the kind

$$\beta = 10^{-4}(a\tilde{\sigma}^2 + b\tilde{\sigma} + c). \tag{4.18a}$$

Here, *a, b, c* are the fitting parameters depending (in a rather complicated manner) on the value of the drag coefficient, $C_d(z)$, and the non-dimensional frequency, $\tilde{\sigma}$, defined as

$$\tilde{\sigma} = \sigma W(\lambda_p)/g \tag{4.18b}$$

where $\lambda_p$ is the wave length corresponding to the peak frequency of the wave spectrum, $\sigma_p$.

The main advantage of the Chalikov's result has two items:

(1) Direct proof of the quadratic power dependence of $\beta$ on W;
(2) Evidence that in the case $W_{10}\sigma/g < 1$, i.e. when waves overcome the local wind, $\beta$ becomes negative with the order of magnitude of $-10^{-5}$.

Disadvantage of numerical simulations is a rather short range of validity, provided by a growing complexity of calculations with an increase of frequency under consideration.

Thus, due to extreme complexity of the problem, the best way is to combine the most reliable experimental and numerical results. Just this way was used by the author (Polnikov, 2005).

First of all, we have accepted the unified empirical result prepared by Yan (1987) from (4.14) and (4.15) in the form

$$\beta_{YAN} = \left\{ \left[ 0.04 \left( \frac{u_* \sigma}{g} \right)^2 + 0.00544 \frac{u_* \sigma}{g} + 0.000055 \right] \cos(\theta - \theta_u) - 0.00031 \right\}. \qquad (4.19a)$$

Secondly, we add a theoretically found and physically justified negative value of $\beta$ for waves running faster than wind. Finally, we have got representation for $\beta$ of the kind

$$\beta = \max\{\beta_L, \beta_{YAN}\}, \qquad (4.19b)$$

where $\beta_L$ is the fitting parameter having the default value

$$\beta_L = 5 \cdot 10^{-6}, \qquad (4.19c)$$

and $\beta_{YAN}$ is given by (4.19a).

Eventually, the most effective and physically justified parameterization of $In$ term has the kind

$$In = C_{in} \sigma \, \beta(\mathbf{k}, \mathbf{W}) \, S(\mathbf{k}, t) \qquad (4.20)$$

with $\beta(\mathbf{k}, \mathbf{W})$ given by formulas (4.19). This parameterization has only two uncertainties: principal fitting coefficient, $C_{in}$, and auxiliary fitting parameter $\beta_L$ having the default value (4.19c).

With respect to previous parameterizations, the proposed parameterization (4.20) has the triple advantage:

(1) Relatively large frequency range of validity, given by rations (4.17);

(2) Negative value of $In$ for waves running faster than local wind;

(3) Rather simple and reasonable mathematical representation.

As a conclusion of the section, and for completeness of our consideration, we could say that in the model WAM they use $In$ term in form (4.20) with $\beta(\mathbf{k}, \mathbf{W})$ corresponding to formula (4.14), and in the model WW they use (4.20) with the Chalikov's result (4.18) for $\beta(\mathbf{k}, \mathbf{W})$. There are a lot of others parameterizations of $\beta(\mathbf{k}, \mathbf{W})$ (see review, The WISE group, 2007), but no of them has any superiority with respect to (4.19).

Our experience of wind wave model constructions ( Polnikov, 1985, 1991) have showed that the approach for $\beta(\mathbf{k}, \mathbf{W})$ used in WAM is too simplified, but that one used in WW is too complicated, with no necessity. Thus, the parameterization of $In$ proposed above in (4.19) is the best compromise of our present understanding the problem of wind-wave energy exchange.

*4.3. Choice of the wind representation*

Finally, we touch a point dealing with the wind representation used in the input term: $W_{10}$, or $u_*$. The principal difference between these too representations is realized only in the case of existing dependence $C_d(z)$ (or $u_*$) on the wave-origin parameters: wind $W_{10}$, wave age $A = c_p / W_{10} = g / \sigma_p W_{10}$, and the other, may be implicit parameters of the wind-waves system. In the case of constant value of $C_d(z)$, there is no principal difference in the wind representation. Thus, the question is the following: can one justify using the complicated form of $C_d(z, W_{10}, A, ...)$ in a practical numerical modelling for wind waves?

In this aspect we note that in such widely used models as WAM and WW, they apply one or another simplified form of dependence $C_d(z)$ on wind $W_{10}$ and wave age $A$, basing on empirical ratios for the former. But there are at least three reasons which give rise doubts of effectiveness of using dependencies of such a kind. They are as follows:

1) As it was shown in Banner and Young (1994), the impact of using a sea-state dependent function for $u_*$ in the input term is negligible for the forecast of wind waves. Very similar result was found in our recent calculations (Polnikov et al., 2002).

2) More over, it is well known (Donelan et al., 1993; Drennan et al., 1999) that empirical data for the drag coefficient, $C_d(z, W_{10}, A, ...)$, and for the sea roughness height, $z_0(W_{10}, A, ...)$, have a very large scattering which provides a great uncertainty in empirical parameterizations for them. The problem of this uncertainty could be solved by constructing a special dynamic boundary layer (DBL) theory permitting to relate the wind-wave state, i.e. local 2D wave spectrum, $S(\mathbf{k})$, with the parameters of the boundary layer of atmosphere, $C_d$, u*, and $z_0$. Our recent study (Polnikov et al, 2003), based on the DBL-theory, proposed in Makin& Kudryavtzev(1999), have shown that all present empirical parameterizations for the dependencies mentioned are not correct. They should be replaced by more accurate (and more complicated) parameterizations, which are not established yet.

3) Finally, the necessity to use theoretically well justified dependencies $C_d(z, W_{10}, A, ...)$ and $z_0(W_{10}, A, ...)$ results in numerous additional calculations, what provides a great time consuming by numerical model, decreasing the practical effectiveness of the forecast.

Nevertheless, the latter circumstance does not mean that one should refuse of calculations for the dependencies said above. But, to our mind, they are rather needed for estimation of the state of atmospheric boundary layer than for the state of wind waves. This point of view is based on the experience of our recent investigations (Polnikov, et al., 2002, 2003) where a preliminary version of a wind wave model with the DBL-block was considered. Just such a kind version of the model, with the DBL-block "switched on", could be classified as the model of new (fourth) generation. There are attempts to make models of such a kind (Janssen, 1991, Polnikov et al, 2002), but the final solution of this problem is waiting its time.

For the completeness of our consideration, we reproduce shortly the main ideas of constructing the DBL-block, which could be "switched on" optionally, if needed in a certain version of a wind wave model.

*4.4. The dynamic boundary block construction*

There are three principal postulates making the basis of the DBL-block construction. They are as follows.

1. One supposes that at the local space point and time moment, the boundary layer system characterized with the constant value of the wind stress

$$\tau = -\rho_a <u'_x u'_z> \equiv \rho_a u_*^2 = const \qquad (4.21)$$

what means that there is no dependence of total stress $\tau$ on vertical coordinate z.

In turn, the total stress, $\tau$, can be shared into two parts:

$$\tau = \tau_t(z) + \tau_w(z) = \rho_a u_*^2 \qquad . \tag{4.22}$$

Here, the first part, $\tau_t$, does not supplies energy to waves, whilst the second part, $\tau_w$, corresponds to the stress which is responsible for the energy pumping of waves. In terms of surface drag (see 4.16), they say that $\tau_t$ is responsible for the "skin" drag, and $\tau_w$ does for the "form" drag ( Donelan, 2003). In Russian literature $\tau_w$ is called as the "wave induced" stress, and $\tau_t$ is called as the "turbulent" or "tangential" stress. Each of these parts can depend on z under the condition of constant value for the total stress, $\tau$.

2. By definition, the wave induced stress can be expresses via wave spectrum by the formula

$$\tau_w(z=0) \equiv \tau_w(0) = \rho_w g \int \frac{k \cos(\theta)}{\omega} In(S,W,\omega,\theta) d\omega d\theta . \tag{4.23}$$

where $In(S,W,\omega,\theta)$ is the well know input term, discussed in subsection 4.2.

3. Dependence of all boundary layer parameters, namely, $C_d$, $u^*$, and $W(z)$, on the sea state can be found from dependence of friction velocity $u_*$ on wave spectrum $S(\mathbf{k})$. To specify these dependencies, one should define dependence $\tau_t$ on $C_d$, $u^*$, $W(z)$, or $S(\mathbf{k})$, and to solve equation (4.22).

Herewith, accepting the logarithmic law for vertical profile of the mean wind

$$W(z) = \frac{u_*}{\kappa} \ln \frac{z}{z_0} \tag{4.24}$$

is the additional condition of this approach.

There are several approaches to construct the DBL-block (Polnikov, 2009b). Non of them are unequivocally acceptable. As an example of a DBL-model which could be a basis of the DBL-block in a wind wave model, we could refer to paper Makin&Kudryavtzev(1999) and give some ideas, following to this paper .

As a particular approach, they accept the following dependence $\tau_t$ on $W(z)$

$$\tau_t(z) = K \frac{\partial W(z)}{\partial z}, \tag{4.25}$$

where $K$ is the vertical mixing coefficient, specification of which was given by a special consideration. Finally, they found

$$W(z) = u_*^2 \int_{z_0^v}^{z} \left[1 - \frac{\tau_w(z)}{u_*^2}\right] K^{-1} dz = \frac{u_*}{\kappa} \int_{z_0^v}^{z} \left[1 - \frac{\tau_w(z)}{u_*^2}\right]^{3/4} d(\ln z) \qquad . \tag{4.26}$$

This principal result permits to specify all other characteristics of the boundary layer.

Namely, after specification of dependence $\tau_w(z)$, they found

$$W(z) = \frac{u_*}{\kappa} \int_{z_0^v}^{z} \left[1 - \frac{J(z)}{1+J(0)}\right]^{3/4} d(\ln z) = \frac{u_*}{\kappa} F(z),$$

where

$$J(z) = \int_{\omega_{min}}^{\omega_{max}} \oint_\theta [\exp(-10zk)\cos(5\pi zk)] \cdot k^2 \cdot \beta(...)S(\omega,\theta)\cos(\theta)d\omega d\theta, \qquad (4.27)$$

and some fitting theoretical parameters are introduced. Equation (4.27) gives the sought boundary layer characteristics:

$$u_* = \kappa W_{10}/F(10), \qquad C_d(10) = (u_*/W_{10})^2 \qquad z_0 = 10/\exp[\kappa U_{10}/u_*]. \qquad (4.28)$$

Here we will not stay more on details of this DBL-model, which has its own advances and shortages. By this consideration we would like to state that a principal way for constriction of the DBL-block is known. Here, it is important to point out state that final solution of this problem permits to solve a lot of additional tasks besides simple wave forecast, basing on a wind wave models of fourth generation (see section 7 below).

## 5. WIND WAVE DISSIPATION MECHANISM

### 5.1. *Overview of the problem*

Let us shortly consider the present state of our knowledge about the dissipation term, *Dis*.

First of all, we should state that there is no any widely recognized spectral representation for function *DIS(S)* (see the most recent review of the state-of-the-art, The WISE group, 2007). In particular, there is not understanding what is the power of spectrum in function *DIS*: the first power (as used in WAM and WW, according to Hasselmann's theory, 1974), the second power (as proposed in Polnikov 1994, 1995, 2005), or the higher power (according to Phillips 1985; Donelan 2001; Hwang&Wang 2004).

Second, there is not any theory where the issue of wind wave dissipation is considered from the most general point of view. All of innumerous theoretical papers, including direct derivations (Hasselmann 1974), dimensional considerations (Phillips, 1985), numerical simulations (Chalikov&Shanin, 1998; Zakharov et al., 2007), all of them do not take into account a presence of small-scale turbulence in the water upper layer, which interacts strongly with waves in reality. They consider, mainly, different aspects of wave breaking processes or accompanying effects, like white-capping in (Hasselmann 1974).

Besides earlier papers by the author (Polnikov, 1994, 1995, 2005), the only exclusion is the paper (Tolman&Chalikov, 1996), where these authors tried to estimate the role of turbulent viscosity in the wave energy dissipation. But they have restricted themselves with a very certain kind of parameterization for the viscosity term, what have led them to numerous, very particular parameterizations vulnerable from a theoretical point of view.

The author's earlier papers mentioned above, dealing with the dissipation of wind wave due to turbulence, are also rather particular and unsystematic. Thus, the problem of construction of more general and logically self consistent theory is strongly in demand at present.

Third, some words about experimental results. Here we will not dwell on review of experimental researches in this field, as far as it is well done in the last paper (Babanin 2007). It needs only to mention, that exhaustive experimental studies of the dissipation mechanism for wind waves are hardly possible, especially in a spectral presentation, due to presence a lot of invisible and immeasurable processes in the water upper layer, governing the dissipation of waves. Therefore, as far as the empirical measurements are restricted by study of different aspects of the wave breaking only, they do not fully correspond to reality.

Nevertheless, for the last 5-10 year, the efforts of experimenters are rather fruitful, and some established empirical effects related to the wind wave breaking process are quite interesting. In particular, following the papers (Banner&Tiang, 1998; Babanin et al., 2001) and some others (see references in Young & Babanin, 2006), we have to mention the following empirical effects related to the wave dissipation process:

E1) Threshold feature of the wave breaking;

E2) Influence of the breaking for long waves on the intensity of breaking for shorter waves;

E3) Different feature of the dissipation rate for dominant waves and for waves in the tail part of wave spectrum;

E4) More intensive breaking of waves running at some angle to the mean wind direction (i.e. two-lobe feature of the angular function for breaking intensity and for wave dissipation rate, consequently).

All of these effects can serve as guidelines, to get a true theoretical representation of *DIS*.

Below we reproduce a short version of the paper (Polnikov, submitted) which is directly devoted to construction the most general and logically self consistent theory of dissipation mechanism for wind wave.

*5.2. Basic statements.*

The main fundamental of the theory states that on the scales of Eq. (2.1) validity, the most general reason for wind wave dissipation is a turbulence of the water upper layer. Herewith, specification of processes producing the turbulence is quite unprincipled.

Really, it is evident that a reasonable part of the turbulence intensity is provided by the wave breaking processes. Thus, the breaking processes are taken into account in our statement, though by an implicit way. Herewith, it is equally clear that accompanying processes, as like as sprinkling, white capping, results of shear currents and wave orbital motions instabilities, taking place both in an atmosphere and in a water boundary layer, bubble clouds, and so on, all of them, in terms of hydrodynamics, are chaotic motions without any determined scale, i.e. the turbulence motions with respect to waves. Contribution of these motions to the wind wave dissipation is to be taken into account, as well. In other words, the proposed theoretical approach is based on including into considerations all dissipative processes leading to production of turbulence in the water upper layer.

According to the said, without any restriction of the consideration generality, the current field in a waving water layer can be written in the form of two constituents

$$\mathbf{u}(\mathbf{x},z,t) = \mathbf{u}_w(\mathbf{x},z,t) + \mathbf{u}'(\mathbf{x},z,t) \ . \tag{5.1}$$

The first summand, $\mathbf{u}_w$, in the right-hand side (further, the r.h.s.) of (5.1), we treat as the potential motion attributed to wind waves. This motion is governed by the system of equations corresponding to one given by Eqs. (2.4)-(2.7), but written in the potential approximation. The second summand, $\mathbf{u}'$, is treated as the turbulent constituent of full velocity, totally uncorrelated with $\mathbf{u}_w$ in statistical sense.

As regards to the surface elevation, $\eta(\mathbf{x},t)$, we do not introduce analogous decomposition, attracting the well known Hasselmann's hypothesis of "a small distortion in mean" for the surface profile, induced by the processes said above (Hasselmann 1974). This allows us to use conception of the surface elevation field, $\eta(\mathbf{x},t)$, in a traditional, commonly used sense.

*5.3. Reynolds stress.*

Following to the said, let us rewrite the basic equations, (2.4) and (2.6), without external force but in the standard tensor kind

$$\frac{\partial u_i}{\partial t} + \sum_j u_j \frac{\partial u_i}{\partial x_j} = -g\delta_{i,3}\big|_{z=\eta(\xi,t)} \quad , \tag{5.2}$$

$$\frac{\partial \eta}{\partial t} = u_3 - \sum_{i=1,2} u_i \frac{\partial \eta}{\partial x_i}\big|_{z=\eta(\xi,t)} \quad . \tag{5.3}$$

Here, the indexes $i, j$ take values 1, 2, 3, corresponding to two horizontal and one vertical coordinates, respectively, and the sub-index, $\big|_{z=\eta(\xi,t)}$, means that the both equations are written at the interface. Further the latter sub-index will be omitted for simplicity.

Substitution of ratio (5.1) into Eqs. (5.2) and (5.3), with the next averaging of them over the turbulent scales and accounting the condition of incompressibility, $\sum_j \partial u_j / \partial x_j = 0$, and absence of a fluctuation part for $\eta(\mathbf{x},t)$, permits us to write the following system of equations valid for the wave motions only

$$\frac{\partial \bar{u}_i}{\partial t} + \sum_j \bar{u}_j \frac{\partial \bar{u}_i}{\partial x_j} = -g\delta_{i,3} - \sum_j \frac{\partial <u_i' u_j'>}{\partial x_j}, \tag{5.4}$$

$$\frac{\partial \bar{\eta}}{\partial t} = \bar{u}_3 - \sum_{i=1,2} \bar{u}_i \frac{\partial \bar{\eta}}{\partial x_i} \quad . \tag{5.5}$$

Here, the mean wave variables, $\bar{\eta}$ and $\bar{u}_i$, are denoted with the bar which will be omitted later with the aim of simplicity, and the additional term in the r. h. s. of Eq. (5.4), appearing due to nonlinearity of the system, is

$$\sum_j \frac{\partial <u_i' u_j'>}{\partial x_j} \equiv P_i \quad . \tag{5.6}$$

Physically, this term represents the disturbing force, **P,** providing for the wave motion dissipation. To convince in this, it is enough to accept a simplest parameterization of the force in terms of wave velocity, **u,** of the kind

$$P_i \equiv \sum_j \frac{\partial <u_i' u_j'>}{\partial x_j} = -\nu_T \sum_j \frac{\partial^2 u_i}{\partial x_j^2} \tag{5.7}$$

with a constant value of factor $\nu_T$. In such a case, term (5.7) is absolutely equal to expression for the typical molecular viscosity force, what, by solution of the system (5.4)-(5.5) in a linear approximation and in a spectral representation for variables $\eta$ and $\mathbf{u}$, leads to the well known equation for the temporal evolution of wave spectrum of the kind (Hasselmann 1960)

$$\partial S(\mathbf{k},t)/\partial t = -2\tilde{\nu}\sigma_k S(\mathbf{k},t), \qquad (\tilde{\nu} = \nu_T k^2 / \sigma_k) \tag{5.8}$$

having an exponentially decay solution (details of Eq. (5.8) derivation can be found in Efimov&Polnikov 1991).

Taking into account that dimension of the introduced constant, $\nu_T$, has one of viscosity, and the solution of (5.8) has a decay feature for the wave spectrum, with no doubt one can state that the additional term (5.6) in Eq. (5.4) has meaning of the dissipative term provided by turbulence in the water upper layer. This simple consideration helps us to draw further very important consequences.

Really, the numerator under the sum in expression (5.6) is a very well known magnitude in the turbulence theory, which is called as the Reynolds stress (Monin&Yaglom, 1971)

$$<u_i^{'} u_j^{'}> \equiv \tau_{ij} \quad , \tag{5.9}$$

which, in our representation, is normalized by the water density. Methods of parameterizations for $\tau_{ij}$ are well developed. Therefore, to complete the theory, it needs to specify a representation for $\tau_{ij}$ in terms of wave variables, $\eta$ and **u**, to find evolution equation of the form (5.8), and to ascribe to the r.h.s. of final equation the physical meaning of dissipation term, *DIS*. To get final result, it is reasonable to use some principles of the Hasselmann's approach to this problem (Hasselmann 1974), while finding the final evolution equation for wave spectrum. Just this program will be realized below.

### *5.4. Phenomenological closure of Reynolds stress*

First of all, let us formulate the main grounds of our concept for a procedure of Reynolds stress closure, the aim of which is to express the turbulent characteristic, $\tau_{ij}$, via the wave variables, $\eta$ and **u**. One of them is the hypotheses of "a small distortion in mean", mentioned above. This hypotheses permits to safe commonly used conception of the wave profile, $\eta(\mathbf{x},t)$, and introduce all derivatives of $\eta(\mathbf{x},t)$, if needed.

The second ground is an assumption that the relative value of Reynolds stress term, $P_i$, is much greater of the dynamical nonlinear term described by the second summand, $\sum_j u_j \frac{\partial u_i}{\partial x_j}$, in the l. h. s. of Eq. (5.4). Thereby, we postulate the statement of "strong" nonlinearity of the turbulent type, permitting to neglect the dynamical nonlinearity in this problem. Thus, we should solve, in fact, the following system of equations

$$\frac{\partial u_i}{\partial t} + g\delta_{i,3} = -P_i(\mathbf{u},\eta) \quad , \tag{5.10}$$

$$\frac{\partial \eta}{\partial t} = u_3 \quad . \tag{5.11}$$

Third. The kind of closure (5.7) is too much simple, reflecting the meaning of "forcing" term (5.6) qualitatively, only. The principle shortage of this closure is reducing the nonlinear dynamics to the linear one. It is evident that more complex, nonlinear closure of the forcing term is more adequate to physics of the processes considered. The problem is to find such a closure which would be more general and have a reasonable physical treatment. Below we try to do it.

One of such a kind version of the stress closure is related to using concept of the Prandtle's mixing length, $L$, allowing to express the turbulent fluctuation velocity, $\mathbf{u}^{'}$, via derivative of the wave velocity field, **u**, in the form (Monin&Yaglom, 1971)

$$u_i^{'} = L_i \partial u_i / \partial x_i \quad . \tag{5.12}$$

In such a case, Reynolds stress becomes a nonlinear function of wave variables

$$\tau_{ij} = <L_i L_i (\partial u_i / \partial x_i)(\partial u_j / \partial x_j)> , \tag{5.13}$$

what changes radically both the structure of final equations and solution of them.

Note that closure (5.12), related to spatial derivatives of the wave velocity, **u**, is quite adequate for a horizontally homogeneous, near-wall turbulence. In such a case, values $L_i$ are ascribed to spatial scales of turbulent eddies, magnitudes of which may be postulated. In our case, the turbulence is realized under conditions of instability of waving interface, therefore, some modifications of the approach are needed. In particular, in the case of waving interface, there are not only possible but needed the closure versions related to derivatives of the wave profile, $\eta(\mathbf{x},t)$, for example, of the kind

$$u_i^{'} = C_i \partial \eta / \partial x_i \tag{5.14}$$

where values $C_i$ (by analogy to $L_\mathbf{i}$) have meaning of scales of mixing velocity.

Another important point, which should be modified in the closure approach, is the averaging procedure over the turbulent scales, applied for the turbulent velocities production, $u_i^{'} u_j^{'}$, to get a proper component of stress $\tau_{ij}$. With the aim to point out this circumstance, we did save the averaging brackets, <.>, in formula (5.13). Meaningfulness of these signs consists in the fact that the wave-like phase structure of those wave variables, $\eta(\mathbf{x},t)$ and $\mathbf{u}(\mathbf{x},z,t)$, which are used for closuring the turbulent value $\tau_{ij}$, should be destroyed (or smoothed) to a certain extent, during the process of averaging over turbulent scales. Here we mean the phase factors alike $\exp[i(\mathbf{k}\mathbf{x})]$, standing under integral in the Fourier-representation of the fields $\eta(\mathbf{x},t)$ and $\mathbf{u}(\mathbf{x}, z, t)$. Such a kind representation is inevitable in the course of constructing a theory for wave spectrum evolution from dynamic equations (5.10)-(5.11) (see below).

The assumption said above is rather extraordinary; nevertheless it has some physical grounds. Really, the wave-like phase information is inappropriate in a turbulent motion. This information should be suppressed to some extent by the averaging procedure over the turbulent scales. Herewith, this reasonable assumption is very fruitful, as far as it allows to manipulate more or less free with the phase factors in the Fourier-representation for the turbulent characteristic, $\tau_{ij}$, and, consequently, for the forcing function, $P_i(\mathbf{u},\eta)$.

On the basis theoretical modifications postulated above, it is quite substantiated to represent the forcing term in the form of rather general quadratic function

$$-P_i(\mathbf{u},\eta) = < \sum_j \frac{\partial}{\partial x_j} \{ [L_i(\partial u_i / \partial x_i) + C_i(\partial \eta / \partial x_i)][L_j(\partial u_j / \partial x_j) + C_j(\partial \eta / \partial x_j)] \} >. \tag{5.15}$$

Taking into account the presumptions done above, we should here emphasize that closure (5.15) maintains the following principal features of the problem:

(a) Nonlinear nature of the dissipation process;

(b) Dependence of the turbulent forcing on gradients of both surface elevation field, $\eta(\mathbf{x},t)$, and velocity one, $\mathbf{u}(\mathbf{x},z,t)$.

Moreover, we have a freedom for manipulation with the phase factors in summand $P_i(\mathbf{u},\eta)$, while making transition to the Fourier-representation for dynamic equations (5.10)-(5.11). All these theoretical grounds have an evident physical meaning.

Besides the physical content, closure (5.15) has an important technical advantage. The latter consists in the fact that the technique of derivation a spectrum evolution equation from dynamic equations (5.10)-(5.11) needs an introduction of generalized Fourier-variable $a_\mathbf{k}$ represented by a linear combination of wave variables $\eta_\mathbf{k}$ and $\Phi_\mathbf{k}$ corresponding to the Fourier–transforms of the elevation and velocity fields (see below). The proposed closure of the kind of (5.15) allows existence a set of stochastic coefficients $L_{i,j}$ and $C_{i,j}$, providing for the Fourier-representation of forcing term, $P_i(a_\mathbf{k})$, in a simple quadratic form of generalized variables. Just this form will be realized below.

The said above allows to state that further specification of coefficients $L_{i,j}$ and $C_{i,j}$ in form (5.15) in not principal at the moment. Moreover, as far as we do not know real processes generating turbulence of the water upper layer, there is no sense to construct any more complicated and detailed approximation for the forcing term, $P_i(\mathbf{u},\eta)$, in the physical space (as they have been done in earlier papers by the author, Polnikov 1993, 1995). At present stage of the theory derivation, it is the most important to take account the nonlinear feature of forcing term only. As it will be shown below, this fact itself gives sufficient grounds for a further finding the general kind of the sought function *DIS(S)*.

Thus, the approach proposed permits to transfer the whole difficulty of choosing specification of the forcing term in a physical space, $P_i(\mathbf{u},\eta)$, to the choice of it in a spectral representation, $P_i(a_\mathbf{k})$.

### 5.5. General kind of the wave dissipation term in a spectral form

Now, return to initial system of equations, (2.4)-(2.7), and rewrite it in the linear and potential approximations without any external force, excluding the turbulent one, $\mathbf{P}(\eta,\mathbf{u})$, introduced in the previous subsection. Accepting the following definitions

$$\mathbf{u}_w(\mathbf{x},z,t) = \vec{\nabla}_3 \varphi(\mathbf{x},z,t) \quad , \tag{5.16}$$

$$\Phi(\mathbf{x},t) \equiv \varphi(\mathbf{x},t)\big|_{z=\eta(\mathbf{x})} \quad , \tag{5.17}$$

one finds that two unknown functions: the surface elevation field, $\eta(\mathbf{x},t)$, and the velocity potential at the surface, $\Phi(\mathbf{x},t)$, are described by the following equations

$$\frac{\partial \Phi}{\partial t} + g\eta = -\hat{P}(\eta,\Phi), \tag{5.18}$$

$$\frac{\partial \eta}{\partial t} = \frac{\partial \Phi}{\partial z}, \tag{5.19}$$

$$\Delta\varphi = 0 \quad \text{and} \quad \frac{\partial \varphi}{\partial z}\bigg|_{z=-\infty} = 0. \tag{5.20}$$

Note that the system (5.19)-(5.20) has the same kind as the system (4.1)-(4.4), except that the last term in the r. h. s. of (5.18) means the result of transition to the potential representation for the turbulence forcing, i.e. $\hat{P}(\eta,\Phi) = (\vec{\nabla}_3)^{-1}[\mathbf{P}(\eta,\mathbf{u})]$. To make a transition into the spectral representation, we introduce, as we done in section 4.1, the following Fourier-decompositions

$$\eta(\mathbf{x},t) = const \cdot \int_\mathbf{k} \exp[i(\mathbf{kx})]\eta_\mathbf{k}(t)d\mathbf{k}, \tag{5.21}$$

$$\varphi(\mathbf{x},z,t) = const \cdot \int_\mathbf{k} \exp[i(\mathbf{kx})]f(z)\varphi_\mathbf{k}(t)d\mathbf{k}. \tag{5.22}$$

After substitution of representations (5.22) into the system of Eqs. (5.18)-(5.20), equations (5.20) give the solution for the potential structure function: $f(z) = \exp(-kz)$, and the other two equations get the kind

$$\dot{\Phi}_\mathbf{k} + g\eta_\mathbf{k} = -\Pi(\mathbf{k},\eta_\mathbf{k},\Phi_\mathbf{k}) \quad , \tag{5.22}$$

$$\dot{\eta}_\mathbf{k} = k\Phi_\mathbf{k} \quad . \tag{5.23}$$

Here, the point above wave variables means the partial derivative in time, and $\Pi(\mathbf{k}, \eta_k, \Phi_k) \equiv F^{-1}[\hat{P}(\eta, \Phi)]$ is the new denotation of forcing function where the operator $F^{-1}$ means the inverse Fourier-transition applied to the forcing function, $\hat{P}(\eta, \Phi)$ (see technical details in Hasselmann, 1974; Polnikov, 2007).

System (5.22)-(5.23) is easily reduced to one equation having a sense of the well know equation for harmonic oscillator with a forcing

$$\ddot{\eta}_k + gk\eta_k = -k\Pi(\mathbf{k}, \eta_k, \dot{\eta}_k) \quad . \tag{5.24}$$

Solution of (5.24), written in the kind of evolution equation for the wave spectrum, can be carried out with the technique used in (Hasselmann 1974).

Following to this technique, introduce the generalized variables

$$a_k^s = 0.5(\eta_k + s\frac{i}{\sigma(k)}\dot{\eta}_k), \quad \text{(where } s = \pm \text{ and } \sigma(k) = (gk)^{1/2}) , \tag{5.25}$$

and rewrite Eq. (5.24) in the kind

$$\dot{a}_k^s + is\sigma(k)a_k^s = -is\sigma(k)\Pi(\mathbf{k}, \eta_k, \dot{\eta}_k)/2g . \tag{5.26}$$

Now, accept the definition of the wave spectrum, used in (Hasselmann 1974)

$$2 << a_k^s a_k^{s'} >> = S(\mathbf{k})\delta(s+s') , \tag{5.27}$$

where the doubled brackets $<<.>>$ mean averaging over the statistical ensemble for wind waves. To finish the evolution equation derivation, one needs to do the following steps:

1) to multiply Eq. (5.26) by the complex conjugated component, $a_k^{-s}$;

2) to sum the newly obtained equation with the original one, (35);

3) to make ensemble averaging the resulting summarized equation.

Finally, one gets the most general evolution equation for wave spectrum of the kind

$$\dot{S}(\mathbf{k}, t) = \frac{2\sigma_k}{g}\text{Im} << \Pi(\mathbf{k}, \eta_k, \dot{\eta}_k)a_k^- >> \equiv -Dis(S) \quad . \tag{5.28}$$

General kind of the sought dissipation term, $Dis(S)$, can be found after specification of the forcing function $\Pi(\mathbf{k}, \eta_k, \dot{\eta}_k)$ based, for example, on the closure formula given by (5.15).

Due to qualitative feature of closure (5.15), there is no need to reproduce here all mathematical procedures explicitly. It is important, only, to take into account the main theoretical grounds providing for the sought final result: the dissipation term as a function of wave spectrum, $Dis(S)$. For more clarity, list below the proper grounds:

(a) The structure of generalized variables (5.25) includes a sum of Fourier-components for elevation variable, $\eta_k$, and for velocity potential one, $\dot{\eta}_k \propto \Phi_k$;

(b) The initial representation of forcing term (5.15) includes analogous sums for derivatives, what means that the forcing term can be expressed via the generalized variables in the form

$$\Pi(\mathbf{k}, \eta_k, \dot{\eta}_k) = function(a_k^s, a_k^{-s}); \tag{5.29}$$

(c) Due to averaging over turbulent scales, the exponential phase factors in the Fourier-representation for $\Pi(\mathbf{k}, \eta_k, \dot{\eta}_k)$ can be arbitrary combined (or simply omitted).

It needs to mention especially that just the item (c) allows executing the inverse Fourier-transitions in the nonlinear summands of forcing term $P(\eta,\Phi)$ without appearance of residual integral-like convolutions containing the resonance-like factors for a set of wave vectors, which are typical in the conservative nonlinear theories (see technical details, for example, in Krasitskii, 1994; Polnikov, 2007). Thus, on basis of the grounds mentioned, it is quite reasonable (and sufficient for the aim posed) to represent the final expression for $\Pi(\mathbf{k},\eta_\mathbf{k},\dot\eta_\mathbf{k})$ in the most simple kind

$$\Pi(\mathbf{k},\eta_\mathbf{k},\dot\eta_\mathbf{k}) = \sum_{s_i,s_j} T_{ij}(\mathbf{k}) a_\mathbf{k}^{s_i} a_\mathbf{k}^{s_j} \ . \qquad (5.30)$$

This form of function $\Pi(\mathbf{k},\eta_\mathbf{k},\dot\eta_\mathbf{k})$ has the main feature of the forcing: nonlinearity in wave amplitudes $a_\mathbf{k}^s$. Herewith, both the explicit kind of multipliers $T_{ij}(\mathbf{k})$ and the certain representation of the quadratic form in the r. h. s. of (5.30) are not principle, as far as the main physical feature is here conserved.

Now, one can get a general kind of the r. h. s. in evolution equation (5.28), using the procedure of multiplication and averaging Eq. (5.26), described above in items 1)-3). First result of this procedure can be found by the following way.

Substitution of (5.30) into (5.28) results in a sum of the third statistical moments of the kind $<<a_\mathbf{k}^{s1} a_\mathbf{k}^{s2} a_\mathbf{k}^{s3}>>$ in the r. h. s. of (5.28). Due to an even power in wave amplitudes for the wave spectrum (by definition (5.27)), any third moment can not be directly expressed via the spectrum function, $S(\mathbf{k})$. In such a case, according to a common technique of the nonlinear theory (see, for example, Krasitskii, 1994; Polnikov, 2007), one should use the main equation (5.26), to write and solve equations for each kind of the third moments, $<<a_\mathbf{k}^{s1} a_\mathbf{k}^{s2} a_\mathbf{k}^{s3}>>$, and to put these solutions into the spectrum evolution equation (5.28).

From the kind of the r. h. s. of Eq. (5.26), it is clear that any third moment will be expressed via a set of the fourth moments of the kind $<<a_\mathbf{k}^{s1} a_\mathbf{k}^{s2} a_\mathbf{k}^{s3} a_\mathbf{k}^{s4}>>$, having a lot of combinations for the superscripts, $s_i$. A part of these moments, for which the condition $s1+s2+s3+s4 \neq 0$ is fulfilled, must be put zero, according to definition (5.27). Residual fourth moments can be split into a sum of products of the second moments, $<<a_\mathbf{k}^{si} a_\mathbf{k}^{-si}>>$, each of which corresponds to the spectrum definition (5.27). By this way, the first nonvanishing summand appears in the r. h. s. of spectrum evolution equation (5.28), and this summand is proportional to the second power in spectrum $S(\mathbf{k})$.

The procedure described can be continued for a part of the fourth moments, what, through the chain of actions described above, results in a sum of terms of the third power in spectrum, in the r. h. s. of evolution equation (5.28). Eventually, the procedure mentioned provides for the power series in spectrum $S(\mathbf{k})$ in the r.h.s. of (5.28), starting from the quadratic term. As far as the whole r. h. s. of Eq. (5.28) has, by origin, a meaning of the dissipative evolution mechanism for a wave spectrum, the proposed theory results in function $DIS(S,\mathbf{k},\mathbf{W})$ of the following general kind[2]:

$$Dis(S,\mathbf{k},\mathbf{W}) = \sum_{n=2}^{N} c_n(\mathbf{k},\mathbf{W}) S^n(\mathbf{k}). \qquad (5.31)$$

---

[2] In a more detailed pose of the problem, instead of simple powers of the spectrum, function $Dis(S)$ could include a set of integral-like convolutions of the same powers in S($\mathbf{k}$). This point is related to a future elaboration of the theory.

Specification of the decomposition coefficients, $c_n$, including their dependence on the wave-origin factors, and determination of the final value of $N$ in series (5.31), is based on principles not related to hydrodynamic equations. Therefore, these points will be specified below, by a separate way.

As a conclusion of this section, it is worth while to emphasize that the main fundamental of the theory, providing for result (40), is nothing else as nonlinear feature of the Reynolds stress closure, substantiated physically in subsection *2.3*. Consequently, the nonlinear feature of result (40) is substantiated at an equal extent.

### 5.6. *Parameterization of the dissipation term and its properties*

In this section, using ideology of the earlier papers (Polnikov 1995, 2005), we will consider the following points:

(a) Certain specification of the dissipation term, *Dis(S ,**k**,**W**)*, of the kind (5.31);

(b) Physical meaning of the parameters introduced;

(c) Correspondence of the parameterization for *Dis(S ,**k**,**W**)* to experimental effects E1-E4 mentioned in subsection 5.1;

(d) Evidence of effectiveness of the proposed version for *Dis(S ,**k**,**W**)*.

<u>5.6.1. Specification of function *Dis*(*S* ,**k**,**W**)</u>

First of all, one should estimate the value of power $N$, which can limit the general representation of *Dis(S ,**k**,**W**)* in the kind of series (5.31). To do it, let us use the following well known fact of existence of a stable and equilibrium spectral shape, $S_{eq}(\sigma)$, usually attributed to a fully developed sea (Komen et al. 1994). Not addressing to discussion about a falling law for the tail part of the wave spectrum, accept here that in the tail part, i.e. under the condition

$$\sigma > 2.5\sigma_p \tag{5.32}$$

($\sigma_p$ is the peak frequency of the spectrum $S(\sigma,\theta)$), the equilibrium spectrum has the shape

$$S_{eq}(\sigma) = \alpha_p g^2 \sigma^{-5} \qquad (\alpha_p \approx 0.01) \tag{5.33}$$

corresponding to the standard Phillips' spectrum (Komen et al, 1994). This assumption gives us a possibility to introduce a small parameter, $\alpha$, defined by the spectral function, $S(\mathbf{k}) \propto S(\sigma,\theta) \propto S(\sigma)$, in the whole frequency band:

$$\alpha = \max[S(\sigma,\theta)\sigma^5 / g^2] << 1 \qquad (0 < \sigma < \infty). \tag{5.34}$$

It is worth while to mention that parameter $\alpha$, defined by (5.34), has the order of the second power of mean wave slope, i.e. it is really quite small ($\alpha \approx 0.01$).

Existence of a small parameter in a spectral representation of the wave field means that any series in spectrum, related to real wave physics, is the series in a small parameter. Hereof, one should immediately conclude that series (5.31) can be restricted by the first term, i.e. $N = 2$, with no lose of theoretical accuracy. Hence, after some algebra, the dissipation function can be written in the form

$$Dis(...) \cong c_2(...)S^2(\mathbf{k}) = \gamma(\sigma,\theta,\mathbf{W})\frac{\sigma^6}{g^2}S^2(\sigma,\theta) \tag{5.35}$$

where the unknown dimensional factor, $c_2$, is changed by the unknown dimensionless function, $\gamma(\sigma,\theta,\mathbf{W})$, all arguments of which a written explicitly. Besides, in the r.h.s. of (5.35), all powers of frequency are brought together to a single one. It is left to define explicit expression for function $\gamma(\sigma,\theta,\mathbf{W})$.

General kind of $\gamma(\sigma,\theta,\mathbf{W})$ can be easily defined on the basis of assumption for existence of equilibrium spectrum, accepted before. By definition, an equilibrium at the tail part of wave spectrum means that for a fixed spectrum shape, $S_{eq}(\sigma,\theta)$, the balance of terms in the source function $F$ is close to zero, i.e.

$$F\big|_{S=S_{eq}} = [Nl + In - Dis]\big|_{S=S_{eq}} \approx 0. \tag{5.36}$$

Now, take into account that the relative contribution of the nonlinear term *NL* to the source function, in the high frequency domain defined by ratio (5.32), is less than 10%. Then, ratio (5.36) gets the form

$$[In - Dis]\big|_{S=S_{eq}} \approx 0. \tag{5.37}$$

Accepting ratio (4.20) as the basic parameterization for the input term, *IN*, by using ratios (5.35) and (5.37) one can easily find a formal expression for function $\gamma(\sigma,\theta,\mathbf{W})$, eventually resulting in the following expression for the dissipation term

$$Dis(\sigma,\theta,S,\mathbf{W})\big|_{S=S_{eq}} \approx \beta(\sigma,\theta,\mathbf{W})\frac{\sigma^6}{g^2}S^2(\sigma,\theta) \quad , \tag{5.38}$$

which is valid in the spectrum tail domain corresponding to ineqality (5.32).

To get final specification of $Dis(\sigma,\theta,S,\mathbf{W})$, valid in the whole frequency domain, one should take into account the following points:

(a) Specific character of dissipation processes in the energy containing domain, i.e. in the vicinity of the spectral peak where $\sigma \approx \sigma_p$;

(b) "Background" dissipation taking place when $\beta(\sigma,\theta,\mathbf{W}) \leq 0$;

(c) Two-lobe feature of the angular spreading function, $T(\sigma,\theta,\theta_w)$, describing an increase of the dissipation rate whilst growing the angular difference, $|\theta - \theta_w|$, i.e. waves do not propagate in the mean wind direction, $\theta_w$.

With account of the said above, finally we have the following specification

$$Dis(\sigma,\theta,S,\mathbf{W}) = c(\sigma,\theta,\sigma_p)\max[\beta_L, \beta(\sigma,\theta,\mathbf{W})]\frac{\sigma^6}{g^2}S^2(\sigma,\theta) \quad . \tag{5.39}$$

Here, the well known increment, $\beta(\sigma,\theta,\mathbf{W})$, is given by a certain empirical formula the kind of which is not principal at the moment (for more details, see Polnikov 2005); $\beta_L$ is the "background" dissipation parameter, the default value of which is $\beta_L = 0.00005$; and $c(\sigma,\theta,\sigma_p)$ is the dimensionless fitting function describing peculiarities of the dissipation rate in the vicinity of spectrum peak frequency, $\sigma_p$. According to (Polnikov 2005), the latter is given by the following phenomenological formula

$$c(\sigma,\theta,\sigma_p) = C_{dis}\max\left[0,\ (1 - c_\sigma(\sigma_p/\sigma)\right]T(\sigma,\theta,\sigma_p) \tag{5.40}$$

where the angular spreading factor is accepted in the form

$$T(\sigma,\theta,\theta_w,\sigma_p) = \left\{1 + 4\frac{\sigma}{\sigma_p}\sin^2(\frac{\theta-\theta_w}{2})\right\}\max\left[1,\ 1-\cos(\theta-\theta_w)\right], \tag{5.41}$$

$C_{dis}$ and $c_\sigma$ are the fitting parameters, and the standard designation, max[$a$, $b$], means a choice of maximum value among two ones under the brackets.

Hereby, the sought parameterization of $Dis(\sigma,\theta,S,\mathbf{W})$ is wholly defined, and a general semi phenomenological theoretical substantiation for the dissipation term is finished. It is left to add that in the course of specification of function $\gamma(\sigma,\theta,\mathbf{W})$, one has a certain arbitrariness related to a choice of the equilibrium spectrum shape, $S_{eq}(\sigma,\theta,\mathbf{W})$, and the kind for the angular form, $T(\sigma,\theta,\theta_w)$, as functions of their arguments. This arbitrariness is justified to some extent by uncertainty of the proper functions obtained experimentally (Rodrigues & Soares, 1999; Young & Babanin, 2006). Nevertheless, the general kind of parameterization (5.35) is robust to the uncertainties of such a kind. Just this circumstance allows hoping on the universality of its application in different numerical models for wind waves.

5.6.2. *Physical meaning of the dissipation term parameters and correspondence to the empirics.* For completeness of the theory, it is important to reveal physical meaning of all innumerous parameters which are used in the proposed version of $Dis(\sigma,\theta,S,\mathbf{W})$, given by formulas (5.39)-(5.41). Without taking into account the fully phenomenological angular spreading function, $T(\sigma,\theta,\theta_w)$, note that the theory has only three fitting parameters: $C_{dis}$, $c_\sigma$ and $\beta_L$.

Meaning of coefficient $C_{dis}$ is evident and simple. It regulates the dissipation intensity. This parameter is inevitable in any approach to the problem of the source function construction. Moreover, just $C_{dis}$ is strongly varied while fitting any numerical model of the kind of (2.1), having representation of the source function as a sum of several separate evolution mechanisms.

Meaning of parameter $c_\sigma$ consists, mainly, in separation of dissipation features in too frequency domains: the vicinity of spectral peak, and the spectrum tail. In our representation, in the vicinity of spectral peak, it regulates an extent of suppression of the "pure turbulent" dissipation intensity described by ratio (5.38). Here one can see manifestation of the empirical effects E2 and E3, mentioned in subsection 5.1.

Really, parameter $c_\sigma$ variation does impact at some extent on the dissipation rate in the spectrum tail domain (a weak analogue of effect E2), what result in the variation of such integral characteristic as the mean wave period, $T_m$. For example, decreasing $c_\sigma$ results in lowering the rate of the spectrum peak growth in the course of wave evolution and, consequently, to increasing a relative contribution of the spectrum tail part to the value of mean period, $T_m$, estimated by the formula

$$T_m = \frac{2\pi\int\sigma^{-1}S(\sigma)d\sigma}{\int S(\sigma)d\sigma}. \tag{5.42}$$

With a glance that a value of the dominant frequency, $\sigma_p$, is mainly defined by the nonlinear mechanism of evolution (Komen et al. 1994, Polnikov 2005, 2009) and weakly depend on the value of $c_\sigma$, the decreasing $c_\sigma$ provides for a mean period reducing, without a remarkable change of the dominant period, $T_p \propto \sigma_p^{-1}$. Naturally, an increase of $c_\sigma$ results in the inverse effect. This feature of parameter $c_\sigma$ was effectively used in the tasks of fitting and verification of new source function (Polnikov at al., 2008; Polnikov&Innocentini, 2008).

Finally, we should say some words about a meaning of parameter $\beta_L$. Its main role is to regulate the dissipation rate at the moments of sharp changing the local wind (falling or turning). It is clear that at these moments, the value of increment $\beta(\sigma,\theta,\mathbf{W})$, corresponding to the former wind direction, is radically reduced, resulting in reducing the rate of breaking for the wave components running along the former wind direction. But in reality, certain (background) turbulence retains, as far as it "lives" in the water upper layer for long time. Thus, the background turbulence should provide remarkable attenuation of the wave components running the former wind direction, which are becoming now a swell. The said explains both a meaning of introducing parameter $\beta_L$ and its role in general. Initial choice of the value for $\beta_L$ is based on numerous empirical and theoretical estimations (see references in Polnikov 2005). But the final value should be found during fitting a whole numerical model, in a concert with a choice of all others parameters.

The joint dynamics of the input term and the dissipation mechanism including a background constituent gives rise a quicker wind sea accommodation to a new wind direction. Just this effect is not described by the numerical models with a traditional dissipation term without a background constituent (WAM or WW), what was explicitly shown in papers (Polnikov at al. 2008, Polnikov & Innocentini 2008) (see below).

Thus, the said above allows stating that all fitting parameters introduced into the proposed version of dissipation term function, $Dis(\sigma,\theta,S,\mathbf{W})$, have both purpose-oriented and physical meaning features.

Additionally, it is worth while to emphasize the important role of quadratic dependence of function *Dis(S)* in the spectrum, *S*. Just this feature allows easy regulating the modeled dependence of the equilibrium spectrum, $S_{eq}(\sigma, \theta)$, on frequency, $\sigma$, by means of varying the frequency power in ratio (5.38). It is caused by the fact that an expression for an equilibrium spectrum shape follows directly from the commonly accepted balance condition (5.37). Really, a substituting the linear in spectrum input term, *IN(S)*, and quadratic in spectrum dissipation term, *DIS(S)*, into Eq. (5.37) gives simply an equation for a shape of spectrum, $S_{eq}(\sigma, \theta)$. Thus, the theory has no restrictions for the shape of equilibrium spectrum.

Particularly, if one wants to postulate the equilibrium spectrum of the Toba's shape (Komen et al 1994)

$$S_{eq,T}(\sigma) = \alpha_T u_* g \sigma^{-4}, \qquad (5.43)$$

he should separately extract the dimensionless multiplier, $g/u_*\sigma$, from function $\gamma(\sigma,\theta,\mathbf{W})$ in (3.35), which should be saved in the r.h.s. of ratios (5.38) and (5.39). In such a case, the balance condition (5.37) results in the sought equilibrium spectrum of the kind (5.43).[3]

A positive comparison of the theoretical version for dissipation term with the well established empirical effects mentioned in subsection 5.1 is finished by the evident fact that the accepted angular spreading function for $Dis(\sigma,\theta,S,\mathbf{W})$ of the kind (5.41) is directly corresponding to the recently revealed empirical fact of two-lobe feature for the angular function discussed (effect E4 found in Young & Babanin 2006). Herewith, regarding to the threshold feature of breaking phenomenon (effect E1), it is easily understood that, in a spectral representation for the dissipation term, this effect is smoothed due to statistical distribution of breaking events in a stochastic wave field alike a wind sea.

---

[3] Note that by this way one changes the dependence of *Dis* on wind W. Thus, the way shown could be used get the best balance between input and dissipation term as functions of the wind (if anybody knows this dependence for equilibrium spectrum).

## 6. VERIFICATION OF NEW SOURCE FUNCTION

In this section will dwell on the problem of estimation the performances of the source function which terms have been discussed previously. This point is directly related to the one for wind wave numerical model as a whole.

There are two approaches to study the numerical model performance: testing and validation processes. The former is based on execution of academic testing tasks, and the latter does on validation of models against natural observation data. In our study, we dealt with the both approaches. As far as the basic principles of these processes have their own specifications, it is worth while to remind them briefly, following to Efimov&Polnikov(1991) and Komen et al. (1994).

### *6.1.    Main regulations for testing and verification of models*

There are three principal features providing for an importance of the testing process. They are as follows:

I. Possibility to reveal numerical features of the model by means of simplified consideration based on using the fully controlled wind and boundary conditions.

II. Message comprehensibility and predictability of the testing tasks.

III. Simple and narrow aimed posing the testing tasks.

There is a long list of testing tasks which could be used for a model properties evaluation (for example, see: The SWAMP group, 1985; Efimov&Polnikov, 1991; Komen et al., 1994; or Polnikov, 2005). But execution of all of them is out of our main aim. At present stage of studying, we have used the following list of tests.

#1. Straight fetch test (the wave development or tuning test).

#2. Swell decay test (the dissipation test).

In general, it is possible to distinguish three levels of adequacy of numerical wind wave models, which are defined by the proper choice of reference parameters used for comparison against observations (Efimov&Polnikov, 1991). But, here we restrict ourselves by the first level only, as far as the checking of the second and third level of adequacy needs much more time and efforts. Example of such a kind testing can be found in Polnikov(2005).

The first level reference parameters are of the most importance, as far they are used in the test #1 which is, in turn, of the to principal importance. They are as follows:

dimensionless wave energy

$$\widetilde{E} = \frac{Eg^2}{W_{10}^4} \qquad \text{(or } E^* = \frac{Eg^2}{u_*^4} \text{)} \qquad (6.1)$$

and dimensionless peak frequency of the wave spectrum

$$\widetilde{\sigma}_p = \frac{\sigma_p W_{10}}{g} \qquad \text{(or } \sigma_p^* = \frac{\sigma_p u_*}{g} \text{)}, \qquad (6.2)$$

where the dimensional energy, $E$, is calculated by the ordinary formula, $E = \iint S(\sigma,\theta)d\sigma d\theta$, and $\sigma_p$ is the peak frequency of the spectrum $S(\sigma,\theta)$. Both values, $\widetilde{E}$ and $\widetilde{\sigma}_p$, estimated from simulations for a stationary stage of the wind wave field, are considered as functions of the non-dimensional fetch,

$$\widetilde{X} = Xg/W_{10}^2. \tag{6.3}$$

Numerical dependences $\widetilde{E}(\widetilde{X})$ and $\widetilde{\sigma}_p(\widetilde{X})$, found in simulations, are to be compared with the reference empirical ratios of the kind (Komen et al, 1994):

a) For the stable atmospheric stratification:

$$\widetilde{E}(\widetilde{X}) = 9.3 \cdot 10^{-7} \widetilde{X}^{0.77}; \qquad \widetilde{\sigma}_p(\widetilde{X}) = 12 \widetilde{X}^{-0.24}, \tag{6.4}$$

b) For the unstable atmospheric stratification:

$$\widetilde{E}(\widetilde{X}) = 5.4 \cdot 10^{-7} \widetilde{X}^{0.94}; \qquad \widetilde{\sigma}_p(\widetilde{X}) = 14 \widetilde{X}^{-0.28} . \tag{6.5}$$

For the test #2, the proper reference parameters are specified below.

On the basis of this comparison, the first tuning of the unknown coefficients in the source function, $C_{in}$, $C_{dis}$, $C_{nl}$, is done. Thus, the sense of these tests is to tune the model. But, here we should to say that the results of this tuning is not an unequivocal (see below), and, in principle, it needs to use more complicated tasks to make a sophisticated tuning. The validation process is one of these tasks.

The only convincing way to prove the superiority of new model (or source function) in solving numerical simulation tasks for wind waves is to carry out the so called procedure of "comparative verification". According to papers (Polnikov at al. 2008, Polnikov & Innocentini 2008), the regulations of comparative verification procedure demand a fulfillment of the following series of conditions:

a) Reasonable data base including accurate and frequent wave observations;

b) Reliable wind field given on a rather fine space-time grid for the whole period of wave observations;

c) Properly elaborated mathematical part for a numerical model of the kind (2.1);

d) Certain numerical wind wave model, chosen for comparison as a reference one.

In papers (Polnikov at al., 2008; Polnikov&Innocentini, 2008), the last two requirements were satisfied by the choice of the model WAM and WW, respectively. Due to very similar results, below we dwell on reproduction the most interesting results of the second paper, only. Just in it, the other conditions, a) and b), were met by the following way.

A. Two oceanic areas were chosen, for which the wave observation data were available: Western and Eastern parts of the North Atlantic (hereafter is referred as NA).

At the first stage of validation process, we have used the one-month data (January, 2006) for 19 buoys located both in the Western and in the Eastern parts of NA (Fig. 5) These data have a time discretization of 1 hour what corresponds to more than 700 points of observations on each buoy.

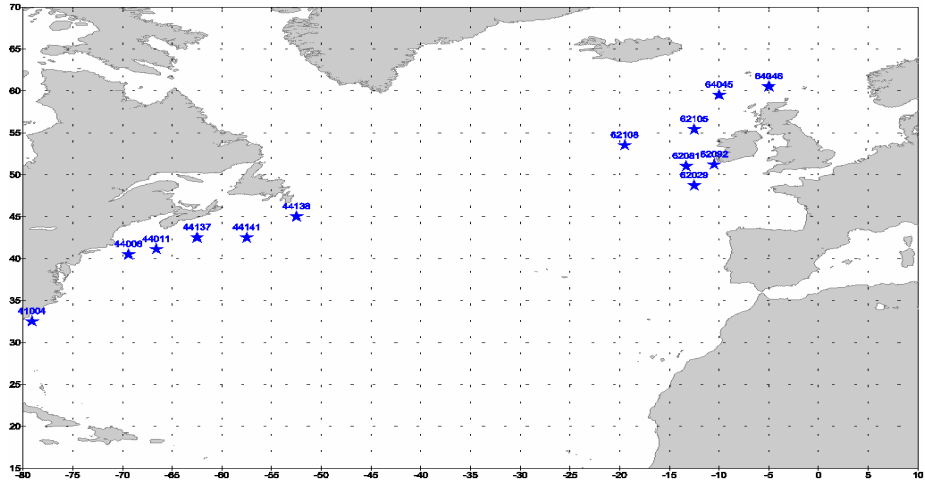

Fig. 5. A sketch of simulating region in the North Atlantic, pointing some buoys locations.

B. As the wind field, we have used a reanalysis made in NCEP/NCAR with a spatial resolution of $1.0^0$ both in longitude and in latitude. The time resolution for the wind was 3 hours. To exclude uncertainties with the boundary conditions, the simulation region was restricted by the following coordinates: $78^0$S – $78^0$N in latitudes and $100^0$W – $20^0$E in longitudes, and the ice covering fields were included into consideration.

The first stage of validation has been executed, basing on the data said above. These calculations resulted in a sophisticated choice of the fitting coefficients, $C_{in}$, $C_{dis}$, $C_{nl}$, found for the default values of the other fitting parameters mentioned above: $b_L$, $\beta_{dis}$, $c_\sigma$ (see below).

At the second stage of validation, we have used the long-period historical data of the National Buoy Data Centrum (NBDC) (covering October-May period of 2005-2006 years) for 12 buoys located in the Western part of NA. The wind fields and the time-space resolution were of the same features as at the first stage. Basing on these data, the standard validation of the both models has been done without changing any coefficients.

As far as the most reliable buoy data are related mainly to the observations of a significant wave height, $H_s$, an estimation of simulation accuracy for this wave characteristic was done in each validation cases. Similar error estimations for a peak wave period, $T_p$, and mean period, $T_m$, were executed for the 2 cases of the long-period simulations with the aim of the work completeness, only. Detailed analysis of the latter estimations is postponed for future investigations.

*6.2. Specification of numerical simulations and error estimations*

In our calculation, we have used the frequency-angle grid of the kind (3.9), having parameters

$$\sigma_0 = 2\pi \cdot 0.04 \text{ rad}, \quad e = 1.1 \quad \text{and} \quad \Delta\theta = \pi/12 \text{ (or } \Delta\theta = 15°) \tag{6.6}$$

with the number of frequency bins $N = 24$ and number of angle bins $M = 24$.

In the case of model testing, the spatial grid was taken in Cartesian coordinates, including 100 points in the *x*-direction and 21 points in the *y*-direction. In the case of model validation in oceanic regions, the grid was taken in spherical coordinates.

The space and time steps of calculations, $\Delta X, \Delta Y, \Delta t$, were varying in accordance with the tasks and the numerical stability conditions. In Cartesian coordinates they varied in limits $\Delta X = \Delta Y = 10^3 - 90 \cdot 10^3$ m  and  $\Delta T = 300 - 900$ s, in spherical coordinates they were $\Delta X = \Delta Y = 1^0$ and $\Delta T = 1200$ s. Every time, an initial spectrum was taken in the frame of WW codes.

To assess an accuracy of simulating a time-series of a certain wave parameter *P(t)*, we have used the following error estimates:

a) The root-mean-square error, $\delta P$, given by the formula

$$\delta P = \left( \frac{1}{N_{obs}} \sum_{n=1}^{N_{obs}} \left( P_{num}(n) - P_{obs}(n) \right)^2 \right)^{1/2} , \qquad (6.7)$$

and

b) The relative root-mean-square error, $\rho P$, defined as

$$\rho P = \left( \frac{1}{N_{obs}} \sum_{n=1}^{N_{obs}} \left( \frac{P_{num}(n) - P_{obs}(n)}{P_{obs}(n)} \right)^2 \right)^{1/2} . \qquad (6.8)$$

Here $N_{obs}$ is the total number of observation points taken into consideration, and the evident sub-indexes are used.

In addition to this, the following arithmetic error was used for analysis:

$$\alpha P = \left( \frac{1}{N_{obs}} \sum_{n=1}^{N_{obs}} \left( P_{num}(n) - P_{obs}(n) \right) \right) . \qquad (6.9)$$

Here we remind that the first two errors describe statistical scattering of the simulating results (or the errors of the input fields, like a wind), whilst the latter one does the mean shift of numerical results with respect to observations.

There are several other statistical characteristics which could be useful for assessment of a numerical model quality (correlation coefficient, probability function, and so on, for example, see Tolman et al, 2002). But at this stage of validation they are omitted, for the sake of more clearness of the primary analysis of the results presented below.

The comparison of error was carried out between numerical results obtained with the original models, WAM and WW, and analogous results done with modifications of the both models, realized by means of replacing the source functions, only. The role of new source function was attributed to one proposed in paper (Polnikov 2005) and described above in sections 3, 4, and 5. This version of modified numerical model (in both cases, WAM and WW) was denoted as the model NEW.

For a further understanding, it is important to note that in both cases, the modification of source functions consists in replacing the terms *Nl* and *In*, represented in the forms (3.10-3.13) and (4.19,4.20), respectively, maintaining the physics enclosed in them.

The modification of *Dis* was done in accordance to the version described by formulas (5.39)-(5.41), changing the physics involved radically. Therefore, the whole deference in accuracy of these calculations was ascribed to changing the term *Dis* in the modified source function. So, in the case of accuracy enhancement, the said permits to say about effectiveness of new version for *Dis*, and vice-verse.

*6.3.  Results of testing new source function*

### 6.3.1. Straight fetch test

Pose of the task. Spatially homogeneous and invariable in time wind, W(x, t) = $W_{10}$= *const*, is blowing normally to a very long straight shore line. Initial conditions are given by a homogeneous wave field with a wave spectrum of small intensity. Boundary conditions are invariable in time and correspond to the initial wave state.

The purpose of the test is to check correspondence of the wind wave growing curves, $\widetilde{E}(\widetilde{X})$, $\widetilde{\sigma}_p(\widetilde{X})$, provided by the model, to the reference empirical growing curves for the stationary state of developed wind waves, given by ratios (6.4), (6.5).

As far as the results of this test are typical and well predicted, here we show only some examples of testing results of the model NEW for different wind values, $W_{10}$ =10-30 m/s. They are presented in Figs. 6-8 for values of $C_{nl} = 9 \cdot 10^7$, $C_{in} = 0.4$, $C_{dis} = 60$, and the default values for the other fitting parameters (see Sects. 3, 4). The proper results for original WW are presented, for example, in Tolman and Chalikov (1996).

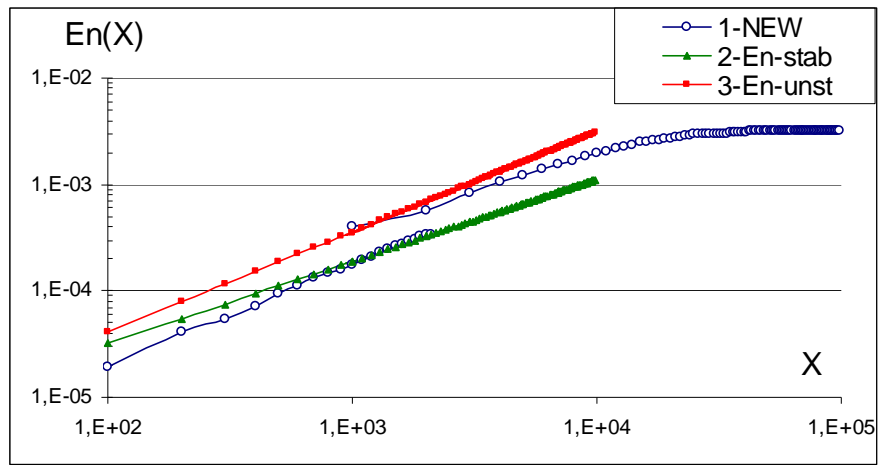

Fig. 6. Dependence of dimensionless energy on dimensionless fetch, $\widetilde{E}(\widetilde{X})$, for $W_{10}$= 10m/s: 1 – model NEW;  2 – Stable stratification ; 3 –  Unstable stratification .

From figures 6-8 one can see that curves $\widetilde{E}(\widetilde{X})$ and $\widetilde{\sigma}_p(\widetilde{X})$ corresponding to the modified model are in a good accordance with empirical ratios (6.4), (6.5). It permits to state a good degree of tuning of the model, securing the first level of its adequacy, at least.

For completeness of treating the results shown, it is worth while to note the following.

First, the jumps between simulation curves, presented in Figs. 3-5, are provided by a change of the spatial step, $\Delta X = \Delta Y$, in 10 times. This change of the spatial step was done in our calculations with the aim to cover a large range of dimensionless fetches, $\widetilde{X}$, for a fixed wind velocity, $W_{10}$.[4] Such a jump for $\widetilde{E}(\widetilde{X})$ and $\widetilde{\sigma}_p(\widetilde{X})$ is a typical feature of any numerical scheme used in the model, consisting in an inevitable dependence of numerical errors on the value of spatial step. Usually, these errors are exaggerated at the points located near the shoreline.[5]

---

[4] For $\Delta X = 10^3$ m and $W_{10}$=10m/s, the range of the non-dimensional fetch $\widetilde{X}$ was $10^2 \leq \widetilde{X} \leq 10^4$, and for $\Delta X = 10^4$ is was $10^3 \leq \widetilde{X} \leq 10^5$.

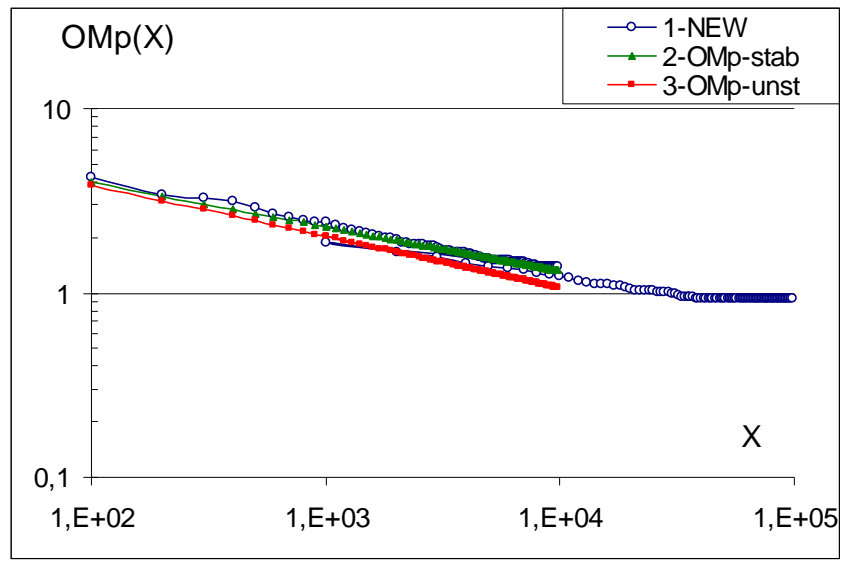

Fig. 7. Dependence of dimensionless peak frequency on dimensionless fetch, $\widetilde{\sigma}_p(\widetilde{X})$.
For legend, see Fig. 6.

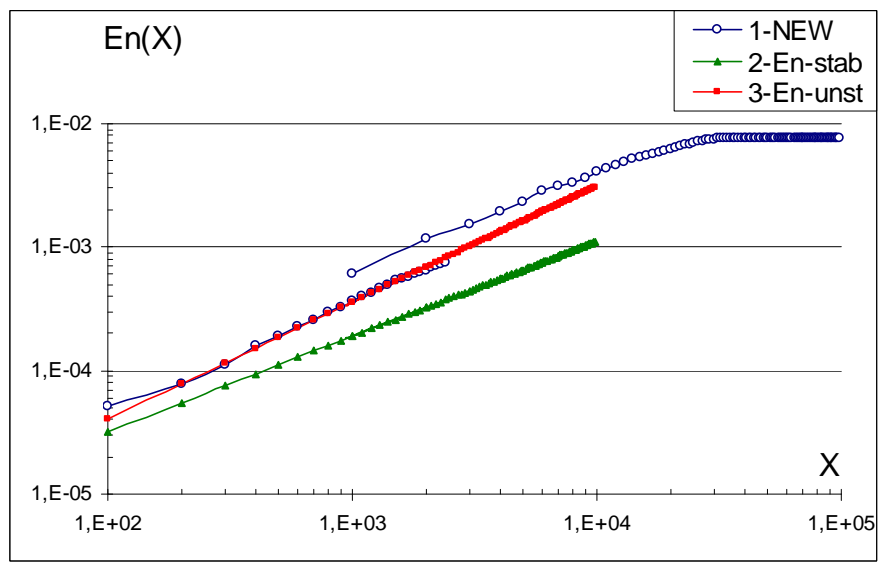

Fig. 8. Dependence of dimensionless energy on dimensionless fetch, $\widetilde{E}(\widetilde{X})$ for $W_{10} = 30$ m/s.
For legend see Fig. 6.

---

[5] This point is mainly related to the mathematical part of the model, which is not discussed here. Evidently that it should be elaborated further in more details.

Additionally, in our presentation of the reference parameters, the location of curve $\widetilde{E}(\widetilde{X})$ is shifted for the same fetch, $\widetilde{X}$, while changing values of $W_{10}$. This result is also typical (see Komen et al, 1994), taking into account dependence of friction velocity, $u_*$, on $W_{10}$, realized in WW. The shifting effect is much more less expressed, if one represents dimensionless parameters in terms of $u_*$ (i.e. the dependencies $E^*(X^*)$ and $\sigma^*_p(X^*)$. But, this artificial effect is not so principal, to dwell on it (for details, one could be referred to Komen et al, 1994; Tolman and Chalikov, 1996).

Second, it should be taken into account that the empirical dependences (6.4), (6.5) are valid for dimensionless fetches of the range $10^2 \leq \widetilde{X} \leq 10^4$ with the errors of the order of 10-15% (Komen et al, 1994). This natural scattering feature of empirical data provides for a possibility to fit a lot of different models to the dependences (6.4), (6.5) with the same accuracy.

Third (and it is of the most importance), a good correspondence of numerical and empirical dependences $\widetilde{E}(\widetilde{X})$, $\widetilde{\sigma}_p(\widetilde{X})$ does not secure an unequivocal choice of the fitting parameters. Coincidence with the root-mean-square errors of the order of 10-15% can be achieved for a continuum of values for the fitting parameters alike of $C_{in}$, $C_{dis}$, $C_{nl}$ and the others mentioned in Sections 2.2-2.4. This result is provided for the simplified meteorological conditions used in the testing task. The sophisticated fitting of the model could be achieved only by means of the model validation against observations executed for a rather long period of wave evolution under well controlled but varying meteorological conditions. This point will be discussed in details below.

### 6.3.2. Swell decay test

Pose of the task. Forcing wind of the fixed values is present in the first part of the testing area: $W(X) = W_{10}$ at points $0 \leq X \leq X_m$. In the second part of the area, the wind is absent: $W(X) = 0$ at $X_m < X \leq 3X_m$. Initial wave state and boundary conditions are typical (see above the pose of test #1).

Numerical evolution is continued for the period $T$ securing a full development of waves at the fetch $X=X_m$ and getting a stable state of the decaying swell field, taking place in the second part of the testing area. Corresponding value of dimensionless time, $\widetilde{T} = Tg/W_{10}$, should be about several units of $10^5$.

The aim of the test is to reveal quantitative features of the swell decay process, starting from the fully developed sea with different peak frequencies, $f_{sw} = f_p(X_m)$. The latter is considered as a principal initial characteristic of the swell. (Here we take into account that the initial intensity of the swell is mainly provided by $f_{sw}$).

To reach the aim posed, different values of $W_{10}$, $X_m$, and $T$ should be taken into consideration. In our calculations, for a force of wind $W_{10} = 10$ m/s, we took: $\Delta X = 10$ km, $X_m = 240$ km, $T = 48$ h; and for $W_{10} = 20$ m/s, we did: $\Delta X = 40$ km, $X_m = 760$ km, $T = 72$ h.

In the second part of the area, the following reference parameters are checked:

- the relative energy lost parameter given by the ration

    $Ren(X) = E(X-X_m)/E(X_m)$; (6.10)

- the relative frequency shift parameter defined as

    $Rf_p(X) = f_p(X-X_m)/f_p(X_m)$. (6.11)

As far as there are no widely recognized empirical dependences *Ren(X)* and *Rf*$_p$*(X)*, the found ones are evaluated at the expert level, only. The latter means a qualitative physical analysis of the numerical results.

Results of our simulation are shown in Figs. 9, 10, representing the swell decay process for values $W_{10}$ = 10 and 20 m/s. The correspondent values of the initial swell frequency, $f_{sw}$, are 0.18 Hz and 0.085Hz, respectively.

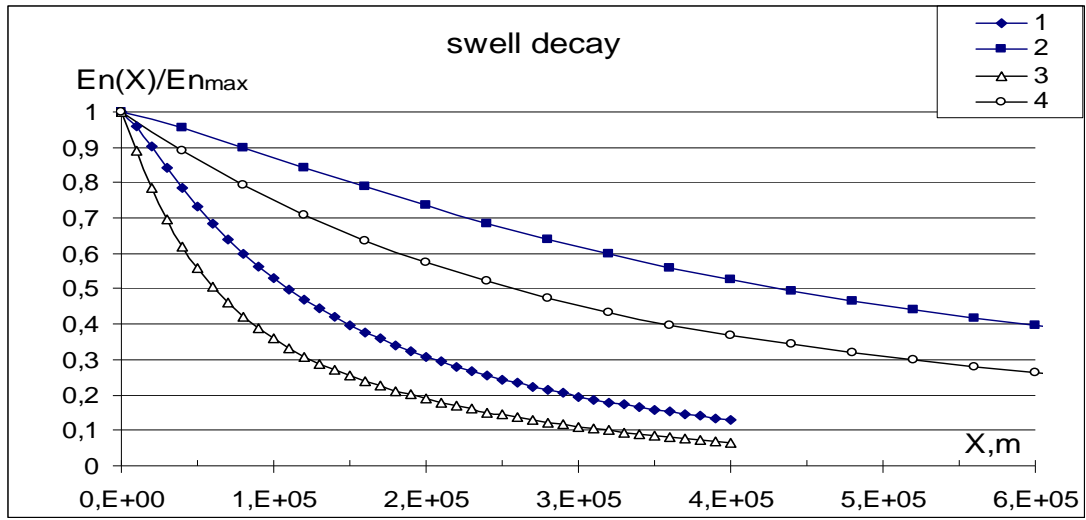

Fig.9. Dependence *Ren*(X) for two values of initial peak frequency of swell:
1, 2 – original model WW; 3, 4 – model NEW; 1, 3 - $f_{sw}$=0.18Hz ; 2, 4 - $f_{sw}$=0.085Hz.

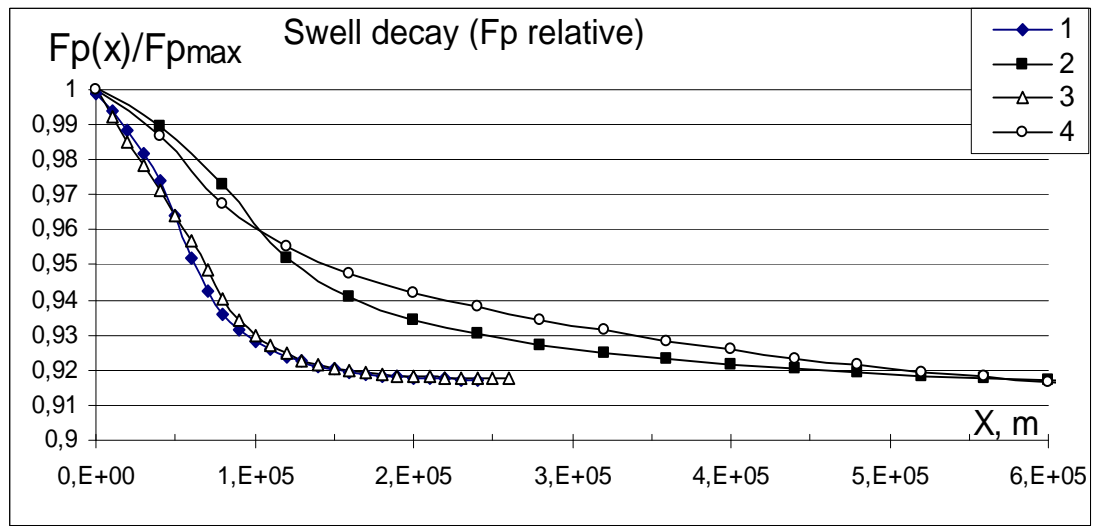

Fig. 10. Dependence *Rf*$_p$(X) for two values of initial peak frequency of swell:
For the legend, see Fig. 9.

From these figures one can draw the following conclusions.

1) The rate of swell energy dissipation depends strongly on the initial peak frequency of swell, $f_{sw}$. This rate is quickly going down with the distance of swell propagation (Fig. 9).

2) The swell dissipation rate for the model NEW is faster than for WW (Fig. 9).

3) The rate of peak frequency shifting to lower values, provided by the nonlinear interaction between waves, depends strongly on the initial value of peak frequency, $f_{sw}$ (Fig. 10). The greater $f_{sw}$ the greater rate of frequency shifting. This is well understood, taking into account formula (3.13) for the nonlinear evolution term.

4) The model NEW has practically the same rate of the peak frequency shifting, in contrast to the rate of the relative energy loss (Fig. 10).

This test is very instructive in the physical aspect. Really, from the results obtained, one can draw the following consequences.

First, from the conclusion 2), one can state that the new dissipation term is more intensive than one used in the original model WW.

Second, from conclusion 4), one can state the fact of very close similarity of the nonlinear terms in the both models.

Third, from previous two consequences, one could state that the main qualitative difference of the numerical results obtained for these two models is mainly provided by the new parameterization of the *Dis*-term. Herewith, we note that though the new parameterization of the *In*-term has a feature of additional background dissipation, in this test it is two small to play any remarkable role, especially at the initial stage of swell decay.

As one could see later, the last consequence is of the most importance for understanding and treatment of difference between these models, which will be found during validation.

### 6.3. *Results of comparative validation of the models WW and NEW*

#### 6.3.1. One-month simulations in the North Atlantic

After several runs of the model NEW, intended to a sophisticated choice of the fitting coefficients $C_{in}$, $C_{dis}$, and $C_{nl}$, we have found that the best results (i.e. minimum errors $\delta H_s$ for the major part of buoys) are obtained for the following values:

$$C_{nl} = 9 \cdot 10^7, \quad C_{in} = 0.4; \quad C_{dis} = 70, \quad \text{and} \quad c_\sigma = 0.7 \tag{6.12}$$

with the default values of the other fitting parameters.

A typical time history of the significant wave height, $H_s(t)$, obtained in these simulations is shown in Fig. 11 for buoy 41001 chosen as an example. From this figure, in particular, one can see that the model NEW follows the extreme values of real waves better than it is done by the model WW. Visual analysis of all proper curves has showed that this feature of the model NEW is typical for the majority of buoys taken into consideration. More detailed and quantitative analysis needs using the statistical procedures based on the error measures described above in Sec 3.4.

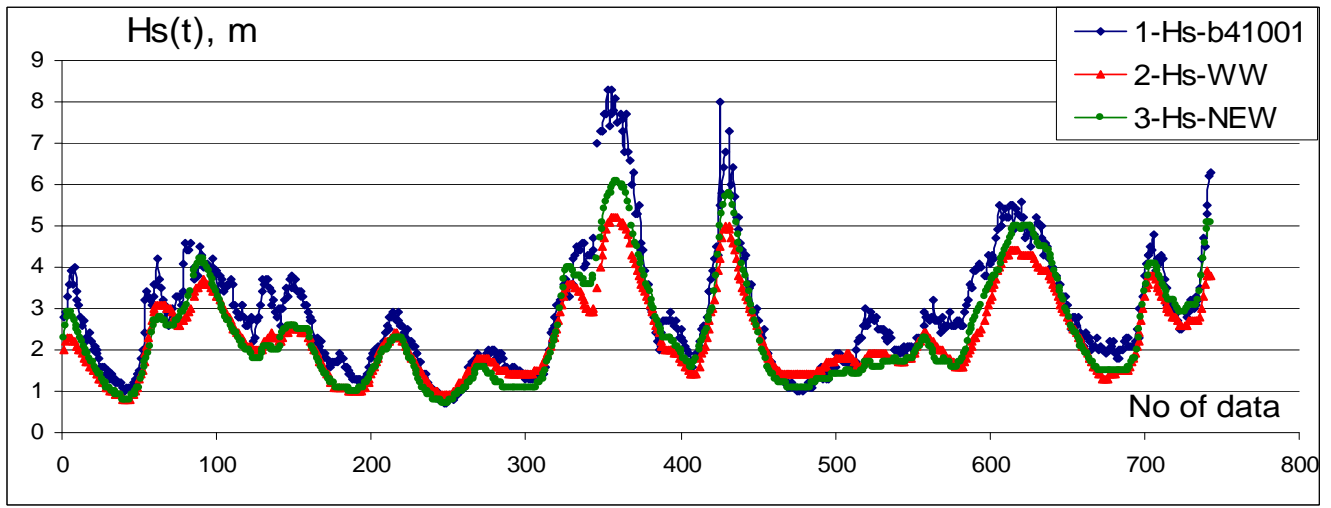

Fig. 11. Time history of the observed and simulated wave heights, $H_s(t)$, on buoy 41001 for January 2006.
1- wave heights measured on the buoy,
2- wave heights simulated by the model WW, 3- wave heights simulated by the model NEW

At this stage of validation, the properly estimated errors have been found for the significant wave height, $H_s$, only. They are presented in Tabs 1, 2 for two parts of NA separately. For a quickness of general (visual) evaluating the results, we have shaded sells corresponding to the cases when the model NEW has a loss of accuracy.

Table 1.

Root-mean-square errors of simulations in the Eastern part of NA

| Eastern NA, | Model WW | | Model NEW | | $\dfrac{(\delta H_s)_{WW}}{(\delta H_s)_{NEW}}$ |
|---|---|---|---|---|---|
| No of buoy | $\delta H_s$, m | $\rho H_s$, % | $\delta H_s$, m | $\rho H_s$, % | |
| 62029 | 0.57 | 14 | 0.54 | 13 | 1.05 |
| 62081 | 0.67 | 15 | 0.56 | 13 | 1.20 |
| 62090 | 0.66 | 14 | 0.57 | 14 | 1.16 |
| 62092 | 0.58 | 14 | 0.53 | 14 | 1.09 |
| 62105 | 0.79 | 18 | 0.68 | 15 | 1.16 |
| 62108 | 0.99 | 15 | 0.84 | 13 | 1.18 |
| 64045 | 0.71 | 12 | 0.61 | 12 | 1.16 |
| 64046 | 0.72 | 15 | 0.76 | 15 | 0.95 |

Analysis of these results leads to the following conclusions.

First, the accuracy of the model NEW is regularly better with respect to one of the original WW. Such a kind result is revealed for more than 70% of buoys considered.

Second, discrepancy of the r.m.s. errors for the both models is remarkable. Typical winning of accuracy for the model NEW is of the order of 15-20%, but sometime it can reach 70% (buoy 44142).

Third, the relative error, $\rho H_s$, calculated by taking into account each point of observations, is not so small (15-27%). It has a tendency of reducing for the model NEW, but this is not so well expressed.

Basing on the said above, we should note that in the present statistical form of consideration, the relative error $\rho H_s$ is not so sensitive to the specificity of the model, as it could be expected. It seems that the effect of more increasing sensitivity of $\rho H_s$ could arise, if we introduce the lower limit of the wave heights, taken into the procedure of error estimation. For example, the proper error estimations could be done, if one excludes the time-series points $H_s(t)$ with the wave values below than 2m. But, a role of introduction of limiting values for $H_s$ (and for $T_p$) is not so evident, therefore this issue should be especially studied later.

Table 2.

Root-mean-square errors of simulations in the Western part of NA

| Western NA | Model WW | | Model NEW | | $\dfrac{(\delta H_s)_{WW}}{(\delta H_s)_{NEW}}$ |
|---|---|---|---|---|---|
| No of buoy | $\delta H_s$, m | $\rho H_s$, % | $\delta H_s$, m | $\rho H_s$, % | |
| 41001 | 0.81 | 22 | 0.66 | 20 | 1.23 |
| 41002 | 0.52 | 18 | 0.47 | 18 | 1.11 |
| 44004 | 0.82 | 25 | 0.68 | 26 | 1.21 |
| 44008 | 0.83 | 27 | 0.61 | 24 | 1.36 |
| 44011 | 0.82 | 23 | 0.55 | 18 | 1.49 |
| 44137 | 0.58 | 19 | 0.51 | 17 | 1.14 |
| 44138 | 0.70 | 19 | 0.74 | 19 | 0.95 |
| 44139 | 0.63 | 19 | 0.69 | 20 | 0.91 |
| 44140 | 0.78 | 19 | 0.80 | 19 | 0.97 |
| 44141 | 0.64 | 20 | 0.68 | 20 | 0.94 |
| 44142 | 0.81 | 27 | 0.48 | 18 | 1.69 |

In this connection, it is worth while to mention about an accuracy of the input wind. The proper time history for $W_{10}(t)$ is shown in Fig. 12.

From the first sight, the correspondence between the simulation wind and the observed wind seems to be rather well. But direct calculations of the errors $\delta W_{10}$ and $\rho W_{10}$, made, for example, for buoy 41001, give the values

$$\delta W_{10} = 1.56 \text{ m/s} \qquad \text{and} \qquad \rho W_{10} = 32\% \,. \tag{6.13}$$

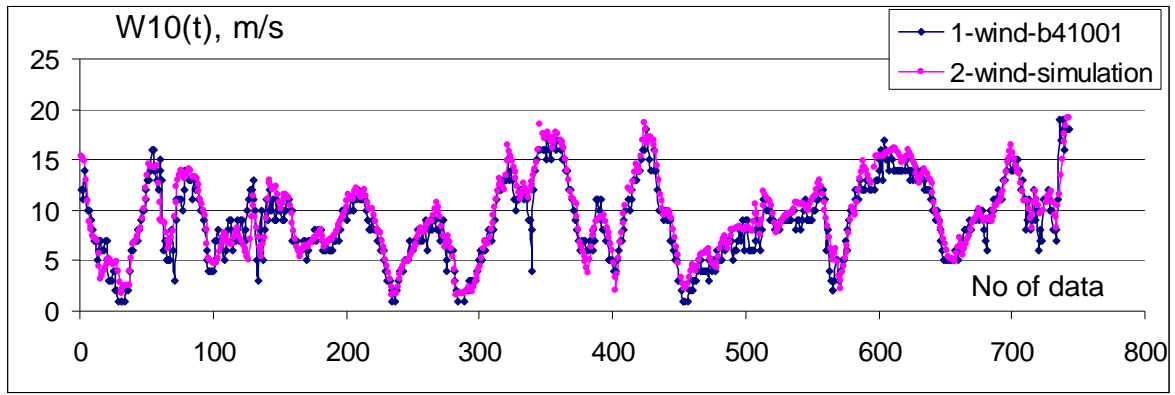

Fig. 12. Time history of the observation and simulation wind, $W_{10}(t)$, on buoy 41001 for January 2006.
1- wind measured on the buoy, 2 – wind used in the modeling simulation

The first value is more or less reasonable, taking into account that the input wind is calculated by the reanalysis for a very large domain covering the whole Earth. But the last value in (6.13), $\rho W_{10}$, seems to be fairly great with respect to the corresponding relative error $\rho H_s$ (Tab. 1). Due to an arbitrary choice of the buoy considered, one can expect that such a kind mismatch between values of $\rho H_s$ and $\rho W_{10}$ is typical for the present consideration, what, in turn, needs its understanding and explanation.

This mismatch of values for $\rho W_{10}$ and $\rho H_s$ leads to a pose of the following new task: how to treat the present inconsistency between these errors. To solve this task, first of all, it needs to have a large statistics of the errors. A part of such a kind statistics will be presented below in Tab. 3. Besides, physically it is reasonable to introduce the lower limiting values for wind, $W_{10}$, and wave heights, $H_s$, which restrict the proper time-series points involved into the procedure of error estimation. In such a way, one could find a physically expected, unequivocal interrelation between errors $\rho W_{10}$ and $\rho H_s$. If it is found, this relation permits to make a proper physical treatment of the errors and to clarify prospective for numerical modeling improvements. Such a kind work is postponed for a future investigation.

6.3.2. Long-period simulations in the Western part of the North Atlantic

Simulating results for the second stage of validation are very similar to the ones presented above. The proper errors are shown the Tab. 3, where the shaded sells correspond to the cases of less accuracy of $H_s$ for the model NEW.

From this table, in general, one can state a reasonable advantage of the new model with respect to WW, in the aspect of simulation accuracy for the wave heights, which is defined by the values of r.m.s. error $\delta H_s$. The winning in accuracy is varying in the limits of 1.1-1.5 times.

More detailed analysis results in the following. Arithmetic errors for WW are regularly greater then ones for the model NEW. Herewith, from table 3 it is seen that the model WW gives permanent underestimation of the wave heights, $H_s$, whilst the model NEW has more symmetrical and smaller arithmetic errors. These facts allows us to conclude that the model NEW (and the new source function, consequently) has apparently better physical grounds.

In the aspect of accuracy for the wave periods, $T_m$ and $T_p$, we should confess that the model NEW has less accuracy of calculation for the mean wave period, $T_m$, but, herewith, it has practically the same (or even better) accuracy for the peak wave period, $T_p$ (Tab. 3).

Regarding to the wave periods, we should note a very specific feature, consisting in the fact that the both models show a certain overestimation of the mean wave period, $T_m$, whilst the peak period, $T_p$, is permanently underestimated. The most probable reason of such a behavior of models could be related to an insufficient accuracy for calculation the 2D-shape of wave spectrum, $S(\sigma, \theta)$, taking place for the both models (for details, see Polniko&Innocentitni, 2008).

Thus, the definite conclusion about superiority of one model against the other can not be drawn at present. Nevertheless, in principle, it could be done later, when the proper criteria will be formulated. This point is only posed here, and we plan to solve it in our future work.

### 6.3.4. Point of the speed of calculation

By using the numerical procedure PROFILE, we have checked the speed of calculation, realized while execution of all main numerical subroutines used in the models. In terms of central processor consuming-time, the proper time distributions among the main subroutines are shown in Tab. 4 for both models. These distributions are corresponding to the case of execution the task of 24-hours simulation of the wave evolution in the whole Atlantic.

From this table one can see that in the model NEW, the nonlinear term is calculated in 1.73 times faster than in the original WW. It leads to the consuming-time winning of the order of 60 seconds, which results in 15%-winning of the total consuming time. The acceleration effect is provided by using the fast DIA approximation mentioned above in Sec. 3. Additional 3%-winning of time is gained due to new parameterization of the input term. But, in turn, the new approximation of *Dis*-term results in a lost of calculation speed on 2%. Nevertheless, as we said above, just this parameterization provides the better accuracy of the model NEW, because of the physics of *NL*-term and *In*-term in both models is very similar.

Table 3.

Consolidated input and output errors for the 8-months simulations in the Western part of NA

| No of buy/model | $\delta W_{10}$, m/s | $\rho W_{10}$, % | $\delta H_s$, m | $\rho H_s$, % | $\delta T_m$, s | $\rho T_m$, % | $\delta T_p$, s | $\rho T_p$, % | $\alpha W_{10}$, m/s | $\alpha H_s$, m | $\alpha T_m$, s | $\alpha T_p$, s | $\dfrac{(\delta H_s)_{ww}}{(\delta H_s)_{new}}$ | $\dfrac{(\alpha H_s)_{ww}}{(\alpha H_s)_{new}}$ |
|---|---|---|---|---|---|---|---|---|---|---|---|---|---|---|
| 41001/WW | 2.01 | 40 | 0.68 | 22 | 0.93 | 17 | 2.02 | 24 | 0.58 | -0.45 | 0.46 | -1.32 | 1.42 | 2.04 |
| /NEW | | | 0.48 | 18 | 1.23 | 22 | 2.13 | 30 | | -0.22 | 0.79 | -0.88 | | |
| 41002/WW | 1.77 | 48 | 0.48 | 19 | 1.20 | 22 | 2.01 | 27 | 0.25 | -0.23 | 0.78 | -1.05 | 1.09 | 7.67 |
| /NEW | | | 0.44 | 20 | 1.58 | 30 | 2.22 | 35 | | -0.3 | 1.11 | -0.57 | | |
| 41004/WW | 2.54 | 36 | 0.97 | 51 | 1.33 | 31 | 2.40 | 36 | -1.48 | -0.97 | 0.63 | -1.26 | 1.52 | 2.06 |
| /NEW | | | 0.64 | 36 | 1.36 | 32 | 2.38 | 38 | | -0.47 | 0.73 | -1.10 | | |
| 41010/WW | 1.24 | 32 | 0.40 | 19 | 1.61 | 33 | 2.06 | 29 | 0.09 | -0.19 | 1.25 | -0.88 | 1.08 | 2.37 |
| /NEW | | | 0.37 | 20 | 2.07 | 43 | 2.34 | 41 | | -0.08 | 1.69 | -0.21 | | |
| 41025/WW | 2.18 | 50 | 0.44 | 24 | 1.47 | 30 | 2.23 | 30 | 0.47 | -0.05 | 1.09 | -1.16 | 0.81 | 0.29 |
| /NEW | | | 0.54 | 31 | 1.82 | 38 | 2.23 | 35 | | 0.17 | 1.48 | -0.57 | | |
| 41040/WW | 0.91 | 20 | 0.22 | 10 | 1.78 | 30 | 1.87 | 18 | 0.08 | -0.10 | 1.64 | -0.90 | 0.88 | 1.43 |
| /NEW | | | 0.25 | 11 | 2.11 | 35 | 1.96 | 22 | | -0.07 | 1.90 | -0.53 | | |
| 41041/WW | 0.96 | 22 | 0.20 | 09 | 1.92 | 32 | 2.22 | 21 | 0.17 | -0.06 | 1.78 | -0.90 | 0.87 | 1.50 |
| /NEW | | | 0.23 | 10 | 2.26 | 38 | 2.20 | 24 | | 0.04 | 2.06 | -0.54 | | |
| 44004/WW | 1.91 | 40 | 0.72 | 24 | 1.13 | 21 | 1.96 | 24 | 0.16 | -0.38 | 0.52 | -1.34 | 1.26 | 9.5 |
| /NEW | | | 0.57 | 24 | 1.32 | 25 | 1.88 | 26 | | -0.04 | 0.82 | -1.00 | | |
| 44005/WW | 2.28 | 59 | 0.58 | 25 | 1.44 | 30 | 2.27 | 38 | 1.19 | -0.30 | 0.84 | -0.84 | 1.29 | 10.0 |
| /NEW | | | 0.45 | 27 | 1.78 | 37 | 2.24 | 43 | | 0.03 | 1.37 | -0.19 | | |
| 44008/WW | 2.35 | 51 | 0.70 | 25 | 1.11 | 21 | 2.01 | 26 | 0.69 | -0.43 | 0.60 | -1.31 | 1.4 | 4.78 |
| /NEW | | | 0.50 | 21 | 1.30 | 25 | 1.88 | 28 | | -0.09 | 0.91 | -0.90 | | |
| 44014/WW | Bad information | | 0.49 | 26 | 1.17 | 21 | 2.46 | 31 | Bad info | -0.27 | 0.46 | -1.66 | 1.4 | 2.45 |
| /NEW | | | 0.35 | 23 | 1.27 | 24 | 2.37 | 33 | | -0.11 | 0.83 | -1.23 | | |
| 44018/WW | 3.01 | 43 | 0.44 | 23 | 1.14 | 22 | 2.03 | 25 | 0.61 | -0.03 | 0.72 | -1.25 | 0.80 | 0.10 |
| /NEW | | | 0.55 | 33 | 1.37 | 28 | 1.88 | 28 | | 0.29 | 1.03 | -0.85 | | |

4Table 4.

Distribution of the central processor consuming-time, realized by the two versions of WW.

| Model | Name of procedure (*explanation*) | Time, s | Time, % |
|---|---|---|---|
| Original WW | w3snl1md_w3snl1 (*Nl-term calculation*) | 123.41 | 27.06 |
| | w3pro3md_w3xyp3 (*space propagation scheme*) | 87.01 | 19.08 |
| | w3uqckmd_w3qck3 (*time evolution scheme-3*) | 68.58 | 15.04 |
| | w3iogomd_w3outg (*output of results*) | 37.73 | 8.27 |
| | w3src2md_w3sin2 (*In-term calculation*) | 21.99 | 4.82 |
| | w3uqckmd_w3qck1 (*time evolution scheme-1*) | 17.66 | 3.87 |
| | w3srcemd_w3srce (*integration subroutine*) | 13.29 | 2.91 |
| | w3src2md_w3sds2 (*Dis-term calculation*) | 2.75 | 0.60 |
| | others | … | … |
| | **All procedures** | **455.9** | **100** |
| Modified WW | w3pro3md_w3xyp3 | 89.72 | 22.52 |
| | w3uqckmd_w3qck3 | 71.29 | 17.88 |
| | w3snl1md_w3snl1 *(Nl-term)* | 70.97 | 17.80 |
| | w3iogomd_w3outg | 38.60 | 9.68 |
| | w3uqckmd_w3qck1 | 17.97 | 4.51 |
| | w3srcemd_w3srce | 12.15 | 3.05 |
| | w3src2md_w3sds2 *(Dis-term)* | 7.68 | 1.93 |
| | w3src2md_w3sin2 *(In-term)* | 6.04 | 1.52 |
| | others | … | … |
| | **All procedures** | **398.8** | **100** |

### 6.3.5. Conclusion for verification

Thus, the new source function was tested and validated by means of its incorporating into the mathematical shell of the reference model WW. Results of the test #1 are typical for any modern numerical model and have a technical feature of the primary tuning. But, the test #2 is more physical, as far as it testifies specific properties of the proposed dissipation term. The real performance of new model was checked during the comparative validation process, which was executed in three steps differing both by duration of simulations and by regions of the World Ocean, taken into consideration.

In general, we may state that the both models have rather high performance, which are apparently the best among present models, taking into account the results of WW's validation represented in Tolman et al. (2002). Herewith, the comparative validation has showed a real advantage of the model NEW with respect to the original WW, especially in the accuracy of the wave heights calculation. The advantage consists in reduction of the simulation errors for significant wave height, $H_s$, in 1.2-1.5 times and increasing the speed of calculation on 15%.

Analysis of the curves, like presented in Fig. 11, shows that the greatest percentage into the r.m.s. error is contributed at the time-series points with extreme values of the wave heights and at the points corresponding to the phases of wave dissipation, at which the wave intensity is going down. Both of these features are controlled by the dissipation mechanism of wave evolution. On these grounds, we conclude that the dissipation term is parameterized more efficiently in the new model than in the original WW. This property of the model NEW is very important in a sense of application it for the tasks of risk assessment.

In our study, the relative r.m.s. error, $\rho H_s$, is introduced, as one of the most instructive measure for estimation an accuracy of the wave heights simulations. We have found that this parameter has mean values of the order of 12-35% for both models. It is naturally to suppose that magnitudes of $\rho H_s$ should be related to the value of inaccuracy for the wind field used as an input. Regarding to this, the new task is posed, consisting in a search for a quantitative relation between the errors for waves, $\rho H_s$, and the errors of input wind, $\rho W_{10}$. This relation is quite expected, taking into account the experimental ratios likes (6.4), (6.5). The proper study is planned to be done in a future work.

There are several another tasks related to the further validation and elaboration of the numerical wind wave models. One of them consists in seeking for a certain upper limits of inaccuracy for wind field and for wave observations, which are requested for a further progress in the wind wave modeling. Estimation of these limits is the primary future task.

At present it seems that the main requirement, which define the limits of the further elaboration of the numerical wind wave models, consists in using the wind field having inaccuracy below the limits mentioned above.

## 7. FUTURE APPLICATIONS

In this section we will point out several possible application of any modern wind wave model, to solve the tasks which are not simply the wave forecast.

### 7.1. A role of wind waves in dynamics of air-sea interface

7.1.1. <u>Introductory words.</u> As we have already said in the INTRODUCTION, wind waves are considered as an intermediate-scale stochastic dynamic process at the air-see interface, which modulates radically the large-scale dynamic processes and small-scale turbulence in the boundary layers of water and air.

Interaction of a large-scale atmospheric air flux (simply say, a wind) with the water upper layer, resulting in the drift currents, is realized at several temporal and spatial scales. Intermediate result of such a kind interaction is the wind sea (wind waves). Here we mean that the large-scale motion of atmosphere and water has variability scales of the order of $10^3$-$10^4$ m and $10^3$-$10^4$ s, whilst the wind wave processes have the scales of the order of 10-$10^2$ m and 10s. In turn, the wave motion of surface interface is realized on the background of small-scale turbulent pulsations of velocity in air and water with characteristics of variability less then 1m and 1s. All these motions characterize dynamics of the air-sea interface. Therefore, the momentum and energy transfer from the wind to currents are realized by means of different direct and inverse cascade transfers in the considered band of temporal and spatial scales.

It is physically evident that the wind waves play a radical modulating role in dynamics of the air-sea interface. But an exact mathematical description of this dynamics meets insuperable difficulties related to multi-scale feature of the processes under consideration. Below we state that mathematical formalization of description for processes of momentum and energy transfer through the air-sea boundary could be done on the basis of well recognized representations of physical mechanisms for wind wave evolution.

Addressing to Fig.1, we distinguish in the air-sea interface 3 constituents:
1) The turbulent atmospheric boundary layer (ABL) with a mean flux of local wind, **W**(**x**), which is characterized by a certain vertical distribution, **W**(**x**, $z$), and a certain value at a fixed standard horizon, for example, by the value **W**$_{10}$(**x**);

2) Waving interface surface, given by the function of wave elevations $\eta$(**x**);

3) Turbulent water upper layer (WUL) with a mean local flux of drift current, **U**(**x**, $z$), distributed by a certain manner through the vertical coordinate $z$.

The vertical dimensions of ABL and WUL are of the order of characteristic (dominant) wave length on the waving interface. The horizontal and temporal scales of variability for all kind of motions under consideration are mentioned above.

One of the main problems of interface dynamics description is the problem of calculation large and scale characteristics of ABL (including wind profile W($z$), and wind stress $\tau$) and characteristics of WUL (including drift current vector **U** and the vertical exchange coefficient K).

It is evident that such a kind detailed description of interface dynamics needs a complicated system of equations, whilst each of them is valid for a certain time-space scale. Attempts of writing such a kind system were undertaken by various authors (for example, see reviews: Monin&Krasitskii, 1985; The WISE group, 2007). In some partial cases, the self-consistent solutions were found (Zaslavskii, 1995; Makin&Kudryavtzev, 1999). But in the general case, a detailed consideration of the problem was not succeeded. An example of such a try is the famous paper by Kitaigorodskii&Lamley 1983.

Usually, in the pose of global circulation tasks, characteristics of wind waves are frequently omitted from the consideration due to small-scale feature of this process (Pedloskii, 1984). However, as it was shown by the practice of theoretical considerations for the wind wave evolution mechanisms, some important particular solutions of the general problem for interface dynamics description, formulated above, can be found just on the basis of well known representations for these mechanisms (Qiao et al, 2004).

From scientific point of view, solution of the circulation problem as a whole is interesting for understanding the general role of dynamics for all items of interface: ABL + wind waves + WUL. On the other hand, from practical point of view, a clear mathematical description of the interface system gives possibility to solve numerous tasks of air-sea dynamics, including diffusion and exchange processes at the interface. Examples of such a kind tasks are as follows:

- self-consistent calculation of wind, waves, and currents;

- calculation of heat, gas, and passive impurity exchange between atmosphere and ocean;

- calculation of mixing in the WUL, including tasks of impurity diffusion and transportation, air bubbles layer generation,

- and so on.

It is easy to see that all of these large and middle-scale processes are in some manner modulated by the wind wave dynamics. By this way the role of wind waves in dynamics of the interface is displayed. The further matching the middle-scale motion with the large-scale one permit to spread a study the wind wave impact on the ocean and atmosphere circulation in a whole.

As the world wide experience shows, attempts of construction a multi-scale equations system in physical variables do not succeed in a full extent, whilst description for dynamics of the system in a spectral representation is more prospective ((Proceedings of AIR-sea interface symposium, 1995, 1999). Herewith, the whole experience gotten in the course of investigation of the wind wave evolution by numerical methods has an essential superiority. Up-to-date, one may state that a principal physical understanding the exchange processes at the air-sea interface is achieved. Numerous results of scientific investigations published, for example, in proceedings of special conferences can be seen as the evidence of this statement (last references). AS we have shown in sections 3, 4, 5 above, the mathematical tool for spectral description of wind waves evolution processes is also well developed. Thus, one may try to construct a series of particular solutions of the general problem formulated above, in the frame of mathematical terms derived already for evolution mechanisms of wind waves.

<u>7.1.2. The role of wind wave evolution mechanisms.</u> First of all, we remind that the waving surface of air-sea interface has a stochastic feature. For this reason, the most adequate description of its motion could be given in terms of two-dimensional wave energy spectrum, $S \equiv S(\sigma, \theta, \mathbf{x}, t)$, spread in space and time and governed by the proper evolution equation (2.1). The heart of this equation is the source function $F$ describing a set of evolution mechanism for wind wave in spectral representation:

- Mechanism of nonlinear energy transfer through the wave spectrum, *Nl* ("nonlinearity-term");

- Mechanism of energy transfer from the wind to waves, *In* ("input-term");

- Mechanism of wave energy loss due to interaction of waves with the turbulence of upper layer, *Dis* ("dissipation-term").

Let us summarize the roles of these mechanisms in the air-sea interface dynamics, i.e. the role of them in distribution of energy between ABL and WUL.

As we have seen, the input mechanism, described by formulas (4.19-4.20) is responsible for energy exchange between ABL and wind waves. The reference wind variable of this process is the friction velocity, $u_*$, the value of which is directly related to the wind stress, $\tau$ (z), corresponding to the vertical turbulent flux of the horizontal momentum, provided by the shear of the wind profile, W(z).

The essential factor of the wind-wave energy exchange process is dependence for both friction velocity and local wind profile, $W(z)$, on the wave state, i.e. on the shape of spectrum $S \equiv S(\sigma, \theta, \mathbf{x}, t)$. Just a determination of this dependence is the task of the feed back influence of wind waves on the state of ABL. In section 4, we have shown that this point can be self-consistently solved by means of construction the model of dynamic boundary layer (DBL). An example of such kind DBL-model was there represented. By this manner, the system of joint description of waves and a boundary layer wind profile becomes closed, and the first part of the problem for interface dynamics, formulated above, does solved.

Secondly, discuss now the role of nonlinear interactions among waves, responsible for the nonlinear evolution mechanism of waves, *Nl*. In section 4, it was shown that nonlinear energy transfer through the wave spectrum has a conservative feature and ensures, mainly, the transfer of wave energy from the high frequency domain of the wave spectrum into the domain located below peak frequency of the spectrum, $\sigma_p$. Such a kind transfer results in diminishing frequency $\sigma_p$ with the time of evolution. It means that during wave evolution due to *Nl*- mechanism, the dominant wave length increasing takes place.

By this manner, the energy, supplied from wind to waves in the high frequency domain mainly, is accumulated by *Nl*-mechanism in the domain below the dominant frequency $\sigma_p$, which is slowly shifting down, whilst the high frequency spectrum tail is practically left unchangeable in intensity (due to balance *F*=0 , in this domain) .

Consequently, just the nonlinear mechanism of evolution procures a significant growth of the total wave energy in the course of their development. Herewith, due to progressive feature of waves, the main part of wave energy is running from the region of its origin into the region of wave propagation. Thus, due to nonlinear feature of wave evolution, the wave energy is not determined by the local wind, only, but it is provided by the whole dynamics of energy exchange during wave propagation through the wave evolution space. Just this fact displays the principal role of term *Nl* in the wave dynamics and, consequently, in the dynamics of energy distributions between items of the interface.

Thirdly, consider the role of wave dissipation mechanism in the air-sea interface dynamics. The direct role of this mechanism is description of the wave energy loss. But the question under consideration is the following: where to and how manner by, the energy lost by waves is distributed?

The answer to the first part of the question is rather clear. The energy lost by waves is shared in two parts. One part of the energy is consumed for generating the WUL turbulence. This part has no any mean momentum, though it has a feature of vertical flux of small-scale turbulent motions, which is spent to the work against buoyancy. The second part of the energy lost by waves is consumed by drift currents. They take a horizontal momentum lost by waves, but it is not the whole momentum taken by waves from wind, due to fact that a part of the whole momentum is gone by waves running to the space of their propagation.

The question of consumption for the energy lost by waves should be considered by the following manner. Estimation of energy partition to the turbulence of WUL and to the drift currents occurrence should be done by the special block of dynamic upper layer (DUL). In such a block, the tasks of joint description for the state of wave, turbulence, and currents in WUL should be solved in a self-consistent manner, in terms of the wave spectrum shape, *S* . In a final form, such a block in not present in modern models till now. Expedience and possibility of its construction is demonstrated by the fact that this point is actively developing last years (see , Ardhuin et al, 2004; Fomin&Cherkesov, 2006).

Thus, the role and importance of adequate representation for all the terms of source function *F* becomes clear and evident.

7.1.3. Energy and momentum balance at the air-sea interface .

The wind is the energy source for all kinds of mechanical motions realized in the items of air-sea interface. Denoting the local large-scale wind at the standard horizon as **W,** let us write expressions for the key characteristics of the wind flux. In particular, a surface density of the energy flux of local wind is

$$\mathbf{F}_{WE} = \rho_a W^2 \mathbf{W}/2 , \qquad (7.1)$$

and local density of the flux of horizontal momentum in given by the ratio

$$\mathbf{F}_{WM} = \rho_a W \mathbf{W} . \tag{7.2}$$

Here $\rho_a$ is the air density and $W$ is the local wind speed. Both these fluxes correspond to the unit of air volume located at the standard horizon.

Due to turbulent feature of the air motion in ABL, the vertical flux of horizontal momentum to the interface, $\tau$, does take place in the system considered. As we said earlier in section 4, $\tau$ can be expressed directly via characteristics of the boundary layer by the formula

$$\tau = \rho_a u_*^2 = \rho_a C_d(z) W^2(z) = \tau_w + \tau_t = const . \tag{7.3}$$

Physically, the energy transfer from wind to waves is just done by the wave part of momentum flux, $\tau_w$, depending on spectrum $S$. On the other hand, the tangent (or turbulent) component, $\tau_t$, give rise to drift current, directly. The task of ABL, wind waves, and UWL dynamics becomes closed in terms of wave spectrum, if one takes into account the circumstance mentioned above, and realizes them as the proper blocks of the generalized wind wave model. Mathematical aspect of this approach is the following.

First, the total value of local density for the wave energy per unit of surface and at the fixed time moment, $E(\mathbf{x},t)$, has the kind

$$E(\mathbf{x},t) = \rho_w g \int_0^\infty d\sigma \int_0^{2\pi} d\theta S(\sigma,\theta,\mathbf{x},t) \quad , \tag{7.4}$$

where $\rho_w$ is the water density, and $g$ is the gravity acceleration.

Second, the rate of energy transfer from wind to waves is unequivocally following from the kind of the input term, *In,* used in the source function of model (2.1). Thus, the total energy flux from wind to waves (per unit of surface) is given by expression

$$I_E = \rho_w g \int_0^\infty d\sigma \int_0^{2\pi} d\theta In[\mathbf{W}, S(\sigma,\theta)] . \tag{7.5}$$

This energy flux is corresponded by a proper momentum flux from wind to waves, which just determines a certain wave component, $\tau_w$, of the total flux, $\tau$. Formula for $\tau_w$ has the kind

$$\tau_w = \rho_w g \int_0^\infty d\sigma \int_0^{2\pi} d\theta \frac{k \cos(\theta - \theta_w)}{\sigma} In[\mathbf{W}, S(\sigma,\theta)] . \tag{7.6}$$

Further construction of the scheme for energy and momentum balance at the air-sea interface, related to the energy transfer from waves into WUL, can be made by analogy with the consideration given above for ABL.

Namely, the rate of energy transfer from waves into the water upper layer is determined by the expression for the dissipation term in the source function of model (2.1), i.e. it has the kind

$$D_E(\mathbf{x},t) = \rho_w g \int_0^\infty d\sigma \int_0^{2\pi} d\theta Dis[\mathbf{W}, S(\sigma,\theta,\mathbf{x},t)] . \tag{7.7}$$

This energy flux is corresponded by the local momentum flux from waves into the WUL

$$\vec{\tau}_d = \rho_w g \int_0^\infty d\sigma \int_0^{2\pi} d\theta \frac{\mathbf{k}}{\sigma} Dis[\mathbf{W}, S(\sigma,\theta)] . \tag{7.8}$$

These two fluxes regulate both the WUL turbulence and drift currents origin. The letter are characterized by the surface density flux of kinetic energy due to currents, $E_C = \rho_w \mathbf{U} U^2 / 2$, and the WUL turbulence do by the rate of turbulent energy production per unit surface square, $E_T$. Ratio between these values is not known in advance, as this needs construction of the dynamic model for WUL (which is hereafter called the block of DUL).

Nevertheless, "*a priori*" one can say that the total momentum flux due to dissipation is shared into the pure turbulent and wave components. Further, with the account of the conservation laws, the energy and momentum fluxes are redistributed among different kinds of motion. Eventually, this mechanics determines the feature of dynamics for the upper mixed layer, which, in the feed back regime, in turn, can modulate the processes of wind wave dissipation. In such a case, the dissipation term *Dis* becomes to be dependent not only on the local wind vector, $\mathbf{W}$, but on the current vector, $\mathbf{U}$, as well.

In full details, the DUL-block construction needs a separate multi-steps consideration. Nevertheless, a whole picture of redistribution of energy and momentum incoming into WUL is rather clear. Herewith, it is evident that just the wind wave dissipation mechanism does play the crucial role in the WUL dynamics, but not the local wind does. The letter is only one of parameters of this mechanism.

Eventually, a general scheme of energy redistribution from the wind to waves, and further into WUL, is presented in Fig. 13.

### 7.2. *Examples for estimation of wave impact on parameters of the ABL and WUL*

It is principally clear that in addition to the wind wave impact on the state of ABL via the energy supply processes, the dissipation processes in waves lead to the origin of turbulent and current motions in the water upper layer. Each of these mechanisms of wave evolution determines radically an intensity of different processes at the interface. In particular, the interface turbulence is responsible to the rate of heat and gas exchange between air and water, to the rate of passive impurity mixing and diffusion in the WUL, and so on. Besides, the wave crests breaking results in an origin of the air bubble layer in the WUL.

Study each the processes mentioned above needs a separate consideration what is out range of this chapter. Therefore, here we present only two examples for quantitative estimations of wave state impact on the parameters of ABL and WUL. One of them will touch the calculation of the friction coefficient, $C_d$, as a function of the wind, inverse wave age, $A$, and the tail shape of 2D-spectrum for wind waves, $S(\sigma, \theta)$. Herewith, the magnitude $A$ is defined by the formula

$$A = u_* \sigma_p / g = u_* / c_p, \tag{7.9}$$

and the spectrum tail shape in a wide range for frequencies with values $\sigma > \sigma_p$, spreading up to the value $\sigma_{max}$ having the order of 80 rad/s, is given by the ratio

$$S(\omega, \theta) \propto \sigma^{-n} \cos^2(\theta - \theta_w). \tag{7.10}$$

The second example deals with the calculation for dependence of the acoustic noise intensity, $I_a$, provided by air bubbles in the WUL, on the local wind speed, $W$.

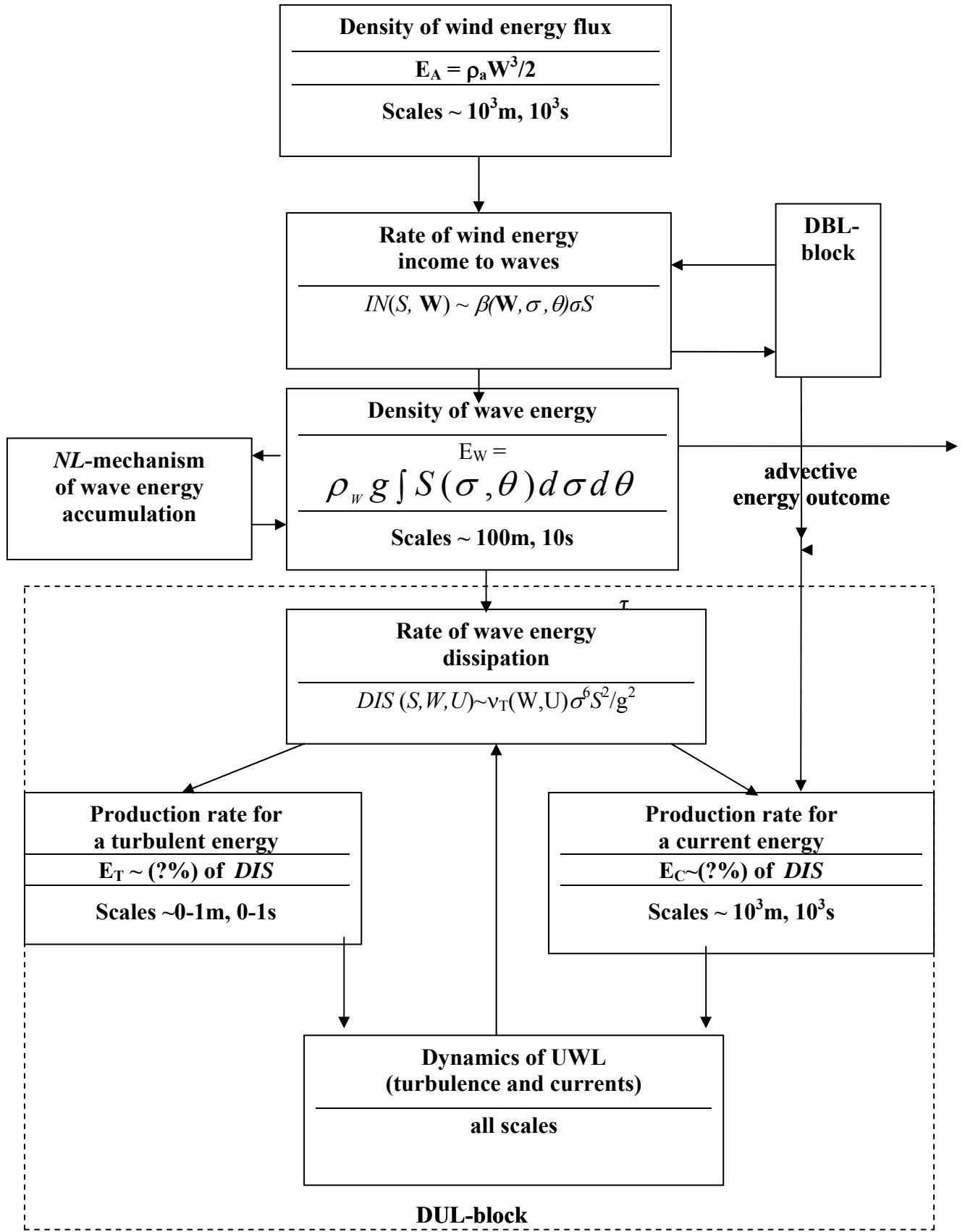

Fig. 13. Scheme of energy redistribution in the air-sea interface

Examples of wave state impact on the drift current in a shallow water basin and for ocean circulation one may find in paper (Fomin&Cherkesov, 2006) and in (Qiao et al., 2004), respectively.

<u>7.2.1. Wave state impact on the value of friction coefficient in the ABL.</u>

This issue was studied in details in paper (Polnikov et al, 2003). First of all, it was noted there that an experimental variability of values for friction coefficient, $C_d$, measured at the horizon z =10m has a dynamical range of variability in the limits of $(0.5-2.5)\times 10^{-3}$ units, for the fixed values of local wind, $W$. Herewith, in the case of swell, the meaning of $C_d$ can get the negative values. The last property of the magnitude $C_d$, as it is clear at present, is totally secured by the inverse energy transfer from waves into the ABL (see section 4). For this reason, below we will not dwell on this point, paying attention on the first point.

For the better understanding the physics of such a kind feature of atmosphere and ocean interaction, the following question should be answered:

- What is the reason of strong variability for values of $C_d$, observable for the same wind speed $W$?

- Is this effect a result of measurements errors or it is provided by physical reasons?

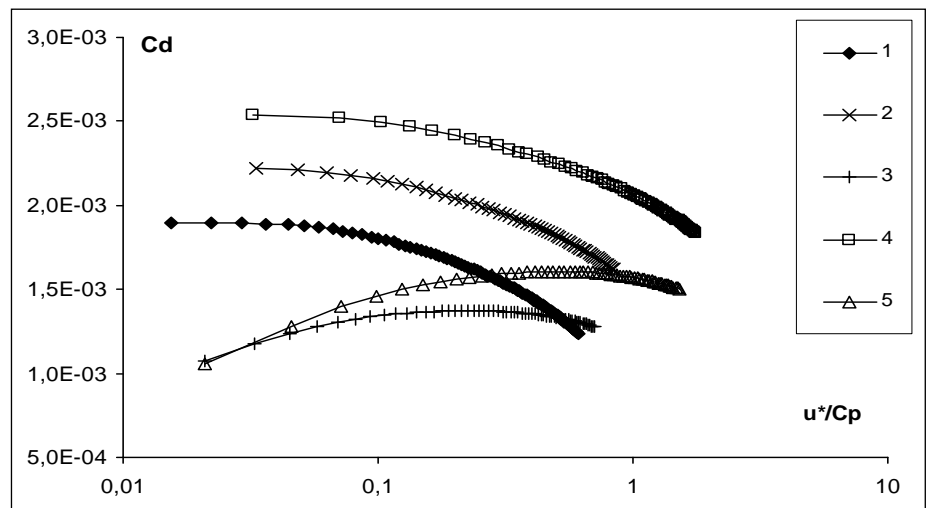

Fig. 14. Dependence of $C_d$ on inverse wave age for series of values for wind, $W$, and spectrum shape parameter, $n$:
1 - $W$ = 5 m/s, $n$ = 4; 2 - $W$ = 10 m/s, $n$ = 4; 3 - $W$ = 10 m/s, $n$ = 5; 4 - $W$ = 20 m/s, $n$ = 4; 5 - $W$ = 20 m/s, $n$ = 5.

The answer to this question was found in (Polnikov et al, 2003) where the calculation of values of $C_d$ were executed with the use of the DBL-model described earlier in subsection 4.4. Results of these calculations for different values of wave age, $A$, and different laws for the spectrum tail fall of the kind of (7.10) are presented in Fig. 14.

Analysis of the results shown in Fig. 14 permits to draw the following conclusions.

1) The scattering of friction coefficient values is secured by the physics of wind-wave interaction process. The value of $C_d$ is determined not only by the local wind speed, $W$, and the current value of wave age, $A$, but by the shape of tail of 2D-spectrum for wind waves, $S(\sigma,\theta)$, too. That explains the wide range for variability of $C_d$, realizing in experimental observations.

2) When the falling laws for the spectrum tail are faster than fifth order in frequency, one may expect the decreasing $C_d$ with decreasing the value of $A$. This effect is frequently observed in experiments (for references, see Polnikov et al, 2003). But for a more weak dependence of the spectrum tail on frequency, it is probable a slow increasing the value of $C_d$ in the course of wave development, which can finish itself by fixing the final value of coefficient, growing with the growth of wind speed, $W$. Consequently, to distinguish the real dependencies $C_d(A, u_*)$, it needs to know (or to calculate) the high frequency tail for 2-D wind wave spectrum, the shape of which is determined by numerous factors including an impact of the surface currents, as well.

3) A simple regression dependence of $C_d$ on parameters of the system considered, alike $W$ and $A$, frequently used in the wave modeling practice (including WAM and WW), is the too rough approximation. Such a kind dependence should be determined by means of the DBL-model, i.e. by means of more elaborated wind wave models (of fourth generation).

Thus, taking into account that the main parameters for a wind profile in the ABL (see formulas 4.28) are calculated with the model of DBL (i.e. without attracting the hypotheses of logarithmic wind profile), one may state that just the presence of the DBL-block in a model leads to appearance a new quality of the latter. This new quality permits to solve the applied tasks for wave forecast and atmosphere circulation with more accuracy and completeness.

To the completeness of consideration, we say several words about redistribution of turbulent and wave components of the vertical momentum flux in the ABL. Numerical estimations of the profile $\tau_w(z)$, made with the DBL-model said above, show that at the mean air-sea surface, the ratio of components $\tau_t$ and $\tau_w$ of the total momentum flux, $\tau$, has a meaning of the following order

$$\frac{\tau_t(0)}{\tau_w(0)} \cong 0.5 - 0.6 \;, \qquad (7.11)$$

where the value z = 0 means a mean level of the waving surface. Such a kind redistribution for the components of $\tau$ leads to a disturbance of the standard logarithmic profile for wind speed, which depends on the wave state. By other words, the use of the standard logarithmic profile for wind speed and estimation of the roughness height $z_0$ with formula (4.28) is the rough approximation to the real situation. Consideration of this issue in more details needs a separate research.

### 7.2.2. Estimation of acoustic noise intensity dependence on the wind speed.

One of important practical task is an estimation of acoustic noise level produced by the air bubbles origin in the WUL due to wave crests breaking. In particular, it is very desirable to know the dependence of bubble noise intensity on the wind speed.

Estimation of relative integral rate of wave energy dissipation, $D_{RE}$, defined by

$$D_{RE} = D_E / E f_p \qquad (7.12)$$

permits to give a theoretical solution of the question posed. Here, $D_E$ is the integral rate of dissipation given by (7.7), $E$ is the total wave energy (7.4), and $f_p$ is the peak frequency. In Polnikov(2009) such estimations were done in a series of simplest cases. It was found that typical value of $D_{RE}$ is of the order of

$$D_{RE} \approx 0{,}001 \quad , \tag{7.13}$$

what leads to the following solution of the problem posed.

Let us rewrite the formulas derived in Tkalich&Chan(2002), where the physical model for acoustic noise of the bubble layer in WUL was constructed. In this paper, it is shown that under certain assumptions, the intensity of bubble noise, $I_a$, is described by the ratio

$$I_a = C_T R(W, H_S) \cdot (f_r / f_0)^{-2} \, , \tag{7.14}$$

where $C_T$ is the theoretical coefficient, $R(W, H_S)$ is the radius of bubble cloud as a function of the local wind speed, $W$, and significant wave height, $H_S$; $(f_0 / f_r)^2$ is the non-dimensional frequency for the bubble acoustic oscillations, depending on a structure of the cloud. Further it is significant only that $I_a$ is linearly dependent on the cloud radius, $R(W)$, the value of which contains a whole information about the wind speed determining the dependence sought.

In Tkalich&Chan(2002), it was shown that the radius value, $R(W, H_S)$, is linearly dependent on the rate of wave energy dissipation in accordance with the ratio

$$R(W, H_S) = c_b c_t \frac{D_E}{Bh} \, . \tag{7.15}$$

Here $D_E$ is the rate of energy income into the WUL due to wave energy dissipation, $c_b \sim (0.3 - 0.5)$ is the empirical coefficient defining the fraction of value for $D_E$ spent to the bubble cloud origin, $c_t \approx 0.5$ is the fraction of $D_E$ spent to the turbulence production in the WUL. $B$ is the void fraction, and $h$ is the characteristic depth of the bubble cloud (mass center). Values of $c_b$ and $B$ are determined from experimental observations, and the value of $c_t$ can be estimated theoretically with the physical models of DBL, discussed above in Sec. 3. Further we will suppose that the values mentioned have a weak dependence on wind. In such a case, the sought dependence takes the kind, $I_a(W) \propto R(W)$, so it can be determined on the basis of calculation for $D_E(W)$ and on physical models describing the dependence $h(W)$.

According to definition (7.12), the dependence $D_E(W)$ can be found from calculations for the magnitude $E(W)$ and from tabulated rate of the non-dimensional dissipation for wave energy, $D_{RE}(W)$. In general case, for the wind field prescribed, this task is solved by means of numerical simulation a wind wave spectrum evolution at the fixed point in the basin under consideration. But in the simplest case of constant and homogeneous wind field, the sought dependences can be obtained with the use of result (7.13) and well known empirical dependences of $\widetilde{E}(W)$ for the fully developed sea, (6.4), (6.5).

Let us consider the fully developed sea. In such a case, with the account of ratios (7.12) and (7.13), one may write

$$D_E(W) = 0.001 \, E_w(W) f_p(W) \, . \tag{7.16}$$

As far as for the fully developed sea, dependences $E_w(W)$ and $f_p(W)$ are given by the well known ratios (6.4), (6.5), we have

$$E_w \approx 3 \cdot 10^{-3} W^4 g^2 \tag{7.17}$$

and

$$f_p \approx g / W 2\pi \, . \tag{7.18}$$

Then, under the assumption of the lack of dependence $B(W)$, for the acoustic noise intensity due to bubbles we have

$$I_a(W) \propto W^3 / h(W). \tag{7.19}$$

Thus, in the case considered, the final result is determined by the model for a depth of the bubbles cloud center, $h(W)$. There are possible the following cases here.

I. In the case of weak wave sea, the assumption that $h = const$ is quite reasonable, due to small dependence of any mechanical parameters for WUL on the wind (including the bubble cloud deepening). In such a case, the dependence $I_a(W)$ is determined by the ratio

$$I_a \sim W^3. \tag{7.20}$$

II. In the case of rather visible waves which are far from their extreme development, it is widely used the following empirical formula (see Tkalich&Chan, 2002)

$$h \approx 0.35 H_S, \tag{7.21}$$

where $H_S$ is the significant wave height. With the account of definition $H_S = 1{,}4(E)^{1/2}$, the sought dependence (7.19) takes the kind

$$I_a \sim W. \tag{7.22}$$

III. And finally, in the case of high winds and fully developed sea, it is more reasonable to put that the bubble cloud depth is linearly related to the radius of the cloud, i.e. $h \propto R$. Under such an assumption and with the account of ratios (7.16)-(7.18), the solution of equation (7.15) reads

$$h(W) \propto W^{3/2}. \tag{7.23}$$

Consequently, in this case, the sought dependence takes the kind

$$I_a \sim W^{3/2}. \tag{7.24}$$

It is interesting to note that all three types of dependences $I_a(W)$, i.e. formulas (7.20), (7.22) and (7.24), do well correspond to generalized observation data presented in Tab. 5, which is

Table 5.

Empirical estimations for dependence $I_a(W)$.

|     | Wind speed (m/s) | Dependence $I_a(W)$ | Wave state |
| --- | --- | --- | --- |
| I   | 5<$W$<10  | $\sim \begin{cases} 0.004W^3 - 0.049W^2 \\ + 0.463W - 1.5 \end{cases}$ | Gravity-capillary and developing waves |
| II  | 10<$W$<15 | | |
| III | $W$>15    | $\sim W^{1.5}$ | Fully developed sea |

taken from paper Tkalich&Chan(2002). It means that the physical assumptions, used above for constructions the models of wind sea and WUL, are fairly adequate to the real processes, and the models themselves can be widely used for solution of practical tasks. Certainly, the solution of acoustic noise problem is still far from its completion. Nevertheless, the main step in this direction has been already done. And it lies in the topic of construction the DBL-model based on the energy and momentum balance established above for the system containing wind, waves, and upper water layer.

### 7.2.3. Intermediate conclusion

Considerations, presented above, permit to look at the whole problem of small-scale and large-scale circulation in atmosphere and upper ocean from the new point of view. Really, up to the present, in the frame of commonly used approximations of geophysical hydrodynamics, solution of the circulation tasks was being executed without account of the waving surface state (Pedloskii, 1984). In such an approach the water surface was considered as an undisturbed one (hard cover approximation), and the momentum transfer from wind into WUL was unequivocally determined by the local wind. On the basis of the said above about the role of wind waves in dynamics of the air-sea interface, the commonly used approach should be radically sophisticated by means of account of the wave state, basing on numerical models including the proper blocks for dynamic ABL and WUL.

A general scheme of the energy and momentum fluxes redistribution in the dynamical system of air-sea interface is presented in Fig. 13. In the frame of this scheme, there is a fairly certain clarity for the wind wave model itself and for DBL-block: there are certain versions of them. Naturally, some details of these models can be sophisticated during their verification, but the principal approach will not get radical changes.

About the model of DBL there is not such a clarity. In particular, there is not any estimation for the fractions of energy and momentum fluxes, going from waving surface to the drift currents and turbulence in the WUL. It needs strong efforts to specify a model of DUL, permitting to close the task of fluxes redistribution and to approach to solution of the circulation task. A series of studies in this direction have been already done (Ardhuin et al., 2004; Fomin& Cherkesov, 2006). And this fact gives a basis to expect an appearance nearest time of new wind wave models installed with the DBL-block.

In relation to the said, it becomes possible to formulate the following sophistication of the present classification for wind wave models, started in (The SWAMP group, 1985).

As it is well recognized in the world practice, the models of the third generation are ones which calculate a full 2-D wind wave spectrum, $S(\sigma,\theta)$, and have source functions operating with no limits for the wave spectrum shape (mainly, this request touches a parameterization for the term *NL*, see, The SWAMP group, 1985). WAM and WW are the most widely used representatives of the models of third generation.

The model of the next generation should have new quality. Such a kind model can be one which is installed with a special DBL-block, permitting to describe a dynamic fitting of the atmospheric boundary layer to the state of wind sea, including calculation of the wind profile without attracting the hypotheses of the logarithmic friction law. The model of such a level can be named as the model of forth generation.

Up to date versions of models WAM and WW, in which the friction velocity is calculated and dynamics of ABL takes place, nevertheless, use the hypotheses of logarithmic ABL and some few-parametric representations for dependences of ABL's parameters on the wave state's ones. Consequently, they do not meet the request formulated for the forth generation model. But the model described in paper (Polnikov, 2005) does meet the request formulated above.

Following to this logics, one may state that the model installed with the block of DUL, permitting to describe dynamics of the water upper layer (the coefficients of turbulent mixing and drift currents, at least), without attraction of few-parametric dependences of them on wind and wave age, will have additional, new quality. Therefore, such a model, installed with the block of DUL, can be classified as the model of fifth generation.

With the account of the said, the classification of wind wave models does get its logical completeness from the theoretical point of view. It is only left to realize the whole chain of models in the practice. A completeness of this chain can be considered as one of the most important theoretical and practical task in the problem of joint description for atmosphere and ocean circulation at the synoptic scales. We suppose that the model based on results of Polnikov (2005) can serve as a basis of the task solution.

*7.3. Using wind wave models for studying long-term mechanical energy exchange in the system: wind-wave-upper ocean*

As an example of alternative application of modern wind wave model, let us consider the following possible project with the conditional title "Wind and wave climate study in Atlantic ocean, based on numerical simulations with a modern model of the third generation".

7.3.1. <u>The main tasks</u>

On the basis of the hind-casting numerical simulations for the 20-years period (1990-2010yy), with the aim of wind and wave climate variability, the following tasks could be done.

1. Calculation and tabulation of the seasonal statistics for wind and wave mechanical energy accumulated in the atmosphere and ocean for 5 regions of Atlantic ocean[6].

2. Annual and seasonal statistics (histograms) of the maximum values of remarkable wave heights (with the threshold $H_s > 3m$) in 5 regions of the Atlantic.

3. Determination of the spatial and temporal distribution of local domains with the extreme wave heights ($H_s > 10m$) in 5 regions of the Atlantic.

4. Annual and seasonal statistics of durations of the extreme waves (with the threshold $H_s > 15$ m) for 5 different regions of the Atlantic.

5. Making electronic maps of wave heights distributions for the extraordinary events in the whole Atlantic (i.e. wind speed is more 30 m/s, or wave heights are of $H_s > 15m$).

7.3.2. <u>Method of study</u>

Having a modern wind wave model (for example, WAM with the optimized source function, as it was done in Polnikov et al, 2008), one could make numerical simulations of wave evolution in the whole Atlantic Ocean for the period of 20 years, to get a good statistics of waves. The proper wind field data are available for us on the space grid $1^0 \times 1^0$ with the time discrete of 3h.

Method of the wave climate study includes the following actions.

A. One makes a spatial partition of the whole Atlantic into 5 parts having, for example, the following boundaries: (X- longitudes, Y – latitudes)

  1. Western part of the North Atlantic (WNA):   100W < X <  40W,   20N < Y < 78N;
  2. Eastern part of the North Atlantic (ENA):     40W  < X < 20E,   20N < Y < 78N;
  3. Tropical part of the Atlantic (TA):           100W < X < 20 E,   20S <  Y < 20N;
  4. Western part of the South Atlantic (WSA):   100W < X <  40W,   78S <  Y < 20S;
  5. Eastern part of the South Atlantic (ESA):    40W  < X < 20E,   78S <  Y < 20S.

B. One introduces 3 reference values of significant wave height, $H_s$, which distinguish description of different meteorological events:

• Ordinary waves heights (with $H_s > 3m$);

---

[6] Partition means fixing the numbers of events in the following 5 regions: Western part of North Atlantic, Eastern part of North Atlantic, Tropical (near-equatorial) part of Atlantic, Western part of Southern Atlantic, and Eastern part of Southern Atlantic.

- Extreme wave heights ($H_s > 10$m);

- Extraordinary wave heights ($H_s > 15$m).

C. In each region of the Atlantics, description of the following events is of interest:

a) Distribution in space and time of the mechanical energy accumulated in atmosphere, $E_A(t)$, (wind) and in ocean, $E_w(t)$, (wind waves) (task 1).

Atmosphere energy time history, $E_A(t)$, is calculated by the formula

$$E_A(t) = \Delta S \sum_{i,j,n} \frac{\rho_a}{2} W_{i,j}^3(t_n) \Delta t_n \tag{7.25}$$

where $\rho_a$ is the air density, and $W_{i,j}(t)$ is the wind at the standard horizon (z = 10m) and at each space-time grid points, (i,j) and time moment $t_n$. Each term under the sum in (7.25) is the density of the kinetic energy flux over a unit of the surface, $\Delta S$.

Wind waves energy analog is calculated by the formula

$$E_w(t) = \Delta S \sum_{i,j} \frac{\rho_w g}{16} H_{i,j}^2(t) \tag{7.26}$$

where $\rho_w$ is the water density. Each term under the sum in (7.26) is the density of the mechanical energy of waves over unit of the surface.

In addition to the said, it is very interesting to calculate a corresponding distribution in space and time of the rate of mechanical energy input into waves from wind, $I_w(R,T)$, and the mechanical energy dissipated by wind waves, $D_w(R,T)$. These values are calculated by the use of the proper source function terms, *In* and *Dis*, applied in the model under consideration.

Proper formulas are as follows

$$I_w(R,T) = \sum_{n \in T} \Delta t \left[ \sum_{i,j \in R} \Delta S \int In(\omega, \theta, W_{i,j}, S_{i,j}) d\omega d\theta \right] \tag{7.27}$$

$$D_w(R,T) = \sum_{n \in T} \Delta t \left[ \sum_{i,j \in R} \Delta S \int Dis(\omega, \theta, W_{i,j}, S_{i,j}) d\omega d\theta \right] \tag{7.28}$$

with the evident sense of notations. Necessity to separate calculation of the mechanical energy input $I_w(R,T)$, in addition to atmospheric and wave energy distributions, $E_A(R,T)$ and $E_w(R,t)$, is provided by the fact that not all the energy $I_w(R,T)$, supplied by the wind to waves, is got, as far as a some part of it is dissipated by waves into the water upper layer.

Calculation of the dissipated energy, $D_w(R,T)$, is interesting to check the balance of the kind

$$B = \{ I_w(R,T) - D_w(R,T) - E_w(R,t) \} \tag{7.29}$$

Positive sing of the magnitude *B* means a presence of a wave energy divergence for the region considered (due to a wave energy advection), but the negative sing does convergence. For the whole world ocean (or a large-scale space) the total balance *B* should be close to zero. One of the points of interest is to check this fact.

Study of these values is important for understanding of the mechanical energy exchange between atmosphere and ocean and their climate variability. 20-years historical series of such values could be needed for estimation of the wave climate variability in time and space.

b) Statistics of the maximum waves. It includes seasonal, annual and total (20-years) histogram of the maximum wave heights, obtained by simulations of wave evolution for 20 years (task 2). This information is important for understanding a regional distribution of wind waves by their strength.

c) Registration of domains with the extreme waves in each region of the Atlantic, and making comparison the numbers of events among the regions (task 3). This information is important for determination of the most dangerous region in the Atlantic.

d) Registration of the extreme wave's duration in the regions (task 3). This information gives more details of the previous study (task 4). It is important for evaluation of the time variability of the extreme events.

e) Making the atlas of maps and seasonal-annual statistics of the extraordinary waves (number of the events in each region) (task 5). This is important for understating of the extraordinary events distribution among regions for the long period. There is no map of such a kind, and for this reason they are of great scientific and practical interest.

The said above does clarify the purpose and the method of executing the project proposed. We sure that the work drafty described in this subsection will be very fruitful in many aspects.